\documentclass[11pt]{article}
\pdfoutput=1
\usepackage{jheppub}
\usepackage{tabu}
\usepackage[vcentermath]{youngtab}
\usepackage[usenames,dvipsnames,table]{xcolor}
\usepackage{graphicx,subfig}
\usepackage{amsmath,amssymb,amsthm,amsfonts,multirow,array,bm,bbm,esint}
\usepackage[mathscr]{eucal}
\usepackage[bbgreekl]{mathbbol}
\usepackage{epsf,grffile}
\usepackage{slashed}
\usepackage[numbers,sort&compress]{natbib}
\usepackage{array,tikz-cd}
\usepackage{braket}
\usepackage{multirow}

\usepackage{subfig}
\usepackage{float}
\definecolor{labelkey}{rgb}{0.4,0.4,0.4}

\newcommand{\ren}{R\'enyi\ }
\newcommand{\ext}{{\rm ext}}
\newcommand{\Tr}{{\rm {Tr}}}
\newcommand{\normord}[1]{:\mathrel{#1}:}
\newcommand{\sech}{\,\mathrm{sech}\,}
 
 \newcommand{\bea}{\begin{eqnarray}}
\newcommand{\eea}{\end{eqnarray}}
\newcommand{\be}{\begin{equation}}
\newcommand{\ee}{\end{equation}}
\newcommand{\ba}{\begin{align}}
\newcommand{\ea}{\end{align}}

\newcommand\rref[1]{(\ref{#1})}

\title{Bootstrapping Quantum Extremal Surfaces I: The Area Operator}

\author[a]{Alexandre Belin,}
\author[b]{Sean Colin-Ellerin}

\affiliation[a]{CERN, Theory Division, 1 Esplanade des Particules, Geneva 23, CH-1211, Switzerland}
\affiliation[b]{Center for Quantum Mathematics and Physics (QMAP),  \\
Department of Physics, University of California, Davis, CA 95616 USA.}

\emailAdd{a.belin@cern.ch, scolinellerin@ucdavis.edu}

\abstract{Quantum extremal surfaces are central to the connection between quantum information theory and quantum gravity and they have played a prominent role in the recent progress on the information paradox. We initiate a program to systematically link these surfaces to the microscopic data of the dual conformal field theory, namely the scaling dimensions of local operators and their OPE coefficients. We consider CFT states obtained by acting on the vacuum with single-trace operators, which are dual to one-particle states of the bulk theory. Focusing on AdS$_3$/CFT$_2$, we compute the CFT entanglement entropy to second order in the large $c$ expansion where quantum extremality becomes important and match it to the expectation value of the bulk area operator. We show that to this order, the Virasoro identity block contributes solely to the area operator.}

\begin{document}

\hfill \\
\begin{flushright}
\hfill{\tt CERN-TH-2021-109}
\end{flushright}

\maketitle

%%%%%%%%%%%%%%%%%%%%%%%%%%%%%%%%%%%%%%%%%%%%%%%%%%%

%~~~~~~~~~~~~~~~~~~~~~~~~~~~~~~~~~~~~~~~~~~~~~~~
\section{Introduction}
\label{sec:Intro}
%~~~~~~~~~~~~~~~~~~~~~~~~~~~~~~~~~~~~~~~~~~~~~~

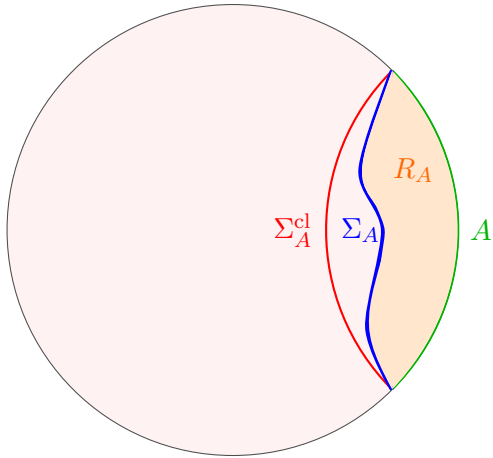
\begin{figure}[h]
\begin{center}
\begin{tikzpicture}
\centering
%AdS3
\filldraw[fill=pink!20!white, draw=black!70!white] (-4.5,0) circle (3cm);
\draw[red,thick] (-2.38,2.12) arc (135:225:3cm);
\node at (-1.2,0) {$\color{green!70!black}A$};
\draw[green!70!black,thick] (-2.38,-2.12) arc (315:405:3cm);
\draw [blue,line width=0.6mm] plot [smooth, tension=0.5] coordinates {(-2.38,-2.12) (-2.71,-1.3) (-2.5,0) (-2.8,0.8) (-2.38,2.12)};
\fill [orange!20!white] (-2.38,2.12) .. controls (-1.22,0.9) and (-1.22,-0.9) .. (-2.38,-2.12);
\fill [orange!20!white] (-2.38,-2.12) .. controls (-2.91,-1.14) and (-2.71,-1.16) .. (-2.15,0.9) -- cycle;
\fill [orange!20!white] (-2.35,0.1) .. controls (-2.95,0.674) and (-2.89,0.676) .. (-2.38,2.12) -- cycle;
\fill [orange!20!white] (-2.35,0.1) .. controls (-2.95,0.674) and (-2.89,0.676) .. (-2.38,2.12) -- cycle;
\fill [orange!20!white] plot [smooth cycle, tension=0.5] coordinates {(-2.6,-1) (-2.52,-0.35) (-2.48,0) (-2.58,0.29) (-2.7,0.53) (-2.2,0.3) (-2.2,-0.3)};
\node at (-3.7,0) {$\color{red}\Sigma_{A}^{\mathrm{cl}}$};
\node at (-2.8,0) {$\color{blue}\Sigma_{A}$};
\node at (-2.1,0.8) {$\color{orange!80!red}R_{A}$};
\end{tikzpicture}
\end{center}
\caption{A fixed time slice of an asymptotically AdS$_3$ spacetime with classical extremal surface $\Sigma_{A}^{\mathrm{cl}}$ (red) given by the spacelike geodesic anchored on the boundary of the CFT interval $A$ (green). The quantum extremal surface $\Sigma_{A}$ (blue) is a spacelike curve that extremizes the generalized entropy and the (quantum) homology surface $R_{A}$ (orange) is the region bounded by $\Sigma_{A}$ and $A$.}
 \label{fig:extremalsurfs}
\end{figure}

The interplay between quantum information theory and quantum gravity has enabled tremendous progress in our understanding of holography and the AdS/CFT correspondence. The star player of this program has been the quantum Hubeny-Rangamani-Ryu-Takayanagi (HRRT) formula, a master formula to compute the entanglement entropy of the boundary CFT semi-classically in the bulk from the generalized entropy of a special surface \cite{Ryu:2006bv,Hubeny:2007xt,Faulkner:2013ana,Engelhardt:2014gca,Jafferis:2015del,Dong:2017xht}
\begin{equation}\label{eq:HRRT}
S_{\textrm{EE}}^{\text{CFT}}(A)= \min\,\underset{\Sigma_A}{\ext}  \left[ \frac{\text{Area}({\Sigma_A})}{4G_N} + S^{\text{bulk}}_{\textrm{EE}}(\Sigma_A) \right] \,,
\end{equation}
where $A$ is a spatial subregion in the CFT and $\Sigma_A$ is a codimension-2 bulk surface that is homologous to $A$. The term $S^{\text{bulk}}_{\textrm{EE}}(\Sigma_{A})$ refers to the entanglement entropy associated to the codimension-1 region $R_{A}$ bounded by $\Sigma_{A}$ and $A$, for all quantum fields that propagate on a given background (see Fig. \ref{fig:extremalsurfs}). The surface $\Sigma_{A}$ that extremizes the generalized entropy in \eqref{eq:HRRT} is called a \textit{quantum extremal surface} and will be the main subject of this work. 

In the context of black hole evaporation, recent developments have established that this master formula is clever enough to compute a Page curve compatible with unitarity, thus taking big steps towards solving Hawking's information paradox \cite{Penington:2019npb,Almheiri:2019psf}. The quantum HRRT formula can ultimately be derived by the semi-classical gravitational path integral \cite{Lewkowycz:2013nqa,Dong:2016hjy,Almheiri:2019qdq,Penington:2019kki,Colin-Ellerin:2020mva}, which appears to be much smarter than previously anticipated. The fact that the semi-classical path integral is capable of reproducing a unitary Page curve suggests that knowing the full microscopic details of the CFT is not crucial at this level. If this is indeed to be the case, it is then natural to ask exactly how much (or what part) of the CFT \textit{is} needed? The use of the Euclidean path integral has unfortunately obscured this aspect and in this work, we will initiate a program to systematically bootstrap quantum extremal surfaces from the microscopic data of the dual conformal field theory, namely a list of operator dimensions and OPE coefficients.

Precisely answering this question in the context of black hole evaporation or for high-energy dynamics is a formidable task, and strong evidence seems to suggest that the gravitational theory has only access to statistical properties of the dual CFT \cite{Pollack:2020gfa,Belin:2020hea,Stanford:2020wkf,Altland:2020ccq,Liu:2020jsv,Belin:2020jxr,Sasieta:2021pzj,Freivogel:2021ivu}. As a starting point, we will consider states that are obtained by acting with low-dimension operators on the CFT vacuum, corresponding to perturbative few-quanta excitations of the quantum fields that propagate in the bulk. For such states, a dictionary between geometric aspects of the quantum extremal surface and the microscopic data of the CFT can be made precise, as we will show.

Very little is known about the properties of quantum extremal surfaces and most explicit computations have been done in AdS$_2$ where the extremal surface is a point (see however \cite{Engelhardt:2019hmr} for general properties based on surface theory). On top of that, it is worthwhile to mention that the formula \eqref{eq:HRRT} suffers from various types of divergences. First, the entanglement entropy in a quantum field theory is UV-divergent due to the entanglement of degrees of freedom close to the entangling surface. This issue is not particularly serious, and one can simply choose to work with CFT quantities that are UV finite such as the relative entropy, the mutual information, or the difference of entanglement entropies between two states. 

On the other hand, the gravitational side of the formula is also bulk-UV divergent. This issue is more subtle and conceptually more involved. The common lore is that the divergence of the bulk entanglement entropy, which should be proportional to the area of the bulk entangling surface, is reabsorbed into Newton's constant which gets renormalized \cite{Susskind:1994sm,Bousso:2015mna} (see also \cite{Belin:2019mlt} for an explicit check in the context of boundary photons and gravitons in AdS$_3$). Running the bulk RG therefore shifts contributions between the area term and the bulk entanglement entropy and it is not a priori clear that one can meaningfully separate the two contributions and attribute either one to some subset of the CFT data. In particular, this also explains why the two terms appear together, since it is really only this combination that the microscopic CFT can ever know about. We will chose to work with quantities that are UV-finite also in the bulk, such that these subtleties do not affect us. We will see that we can unambiguously attribute CFT contributions to either term.

%~~~~~~~~~~~~~~~~~~~~~~~~~~~~~~~~~~~~~~~~~~~~~~~
\subsubsection*{Pertubative states}
\label{sec:Pertstates}
%~~~~~~~~~~~~~~~~~~~~~~~~~~~~~~~~~~~~~~~~~~~~~~

The simplest state we can think of is the vacuum of the CFT. Unfortunately, the entanglement entropy of a single region in the vacuum of a CFT is fixed by symmetry (at least for sufficiently symmetric entangling surfaces), and the only effect of quantum extremality is to renormalize Newton's constant. To probe the dynamics of the theory, one must either consider multiple intervals (see for example \cite{Headrick:2010zt,Hartman:2013mia,Faulkner:2013yia,Barrella:2013wja,Headrick:2015gba,Belin:2017nze}) or change the state, as we will do here.

A particularly nice class of states to study are perturbative states, corresponding to the excitation of a few quanta of the perturbative bulk fields that propagate on the AdS vacuum. It is expected that such states can be treated purely within the bulk low energy effective field theory. The simplest among such states correspond to one-particle states of the bulk fields \cite{Belin:2018juv}, which in the CFT maps to the insertion of a single-trace primary operator at the origin
\begin{equation}\label{eq:bulkstateCFTstate}
\ket{\psi}_{\text{bulk} }= a^{\dagger}_{(0,0)} \ket{0}_{\text{global AdS}}  \qquad \Longleftrightarrow \qquad \ket{\psi}_{\text{CFT} } = O(0) \ket{0}_{\text{CFT}} \,.
\end{equation}
Such states provide nice testing grounds for the entanglement entropy where the answer is not fixed by symmetry and depends dynamically on the CFT data. It is important to emphasize that these are truly quantum states of the bulk theory which are not dual to (semi)-classical geometries, unlike coherent states prepared by a Euclidean-path integral with sources \cite{Botta-Cantcheff:2015sav,Marolf:2017kvq,Belin:2018fxe,Belin:2020zjb}, or heavy operators dual to black hole microstates \cite{Fitzpatrick:2014vua,Asplund:2014coa}. In particular, order by order in $G_N$, changes in the geometry due to quantum backreaction are of the same order as the change in the bulk entanglement entropy.

%~~~~~~~~~~~~~~~~~~~~~~~~~~~~~~~~~~~~~~~~~~~~~~~
\subsection{General program: Bootstrapping quantum extremal surfaces}
\label{sec:Genpro}
%~~~~~~~~~~~~~~~~~~~~~~~~~~~~~~~~~~~~~~~~~~~~~~

This paper is the first in a series of papers aimed at constructing a dictionary between CFT OPE data given by the conformal dimensions and OPE coefficients of single- and double-trace operators, and the contributions to the expectation value of the area operator and the bulk entanglement entropy appearing in the generalized entropy. 

For convenience, we will work in AdS$_3$/CFT$_2$.\footnote{In higher-dimensions, we expect a generalization of our results using the technology of \cite{Sarosi:2016atx} with a suitable treatment of graviton entanglement on the bulk side.} In 2d CFTs, the difference of \ren entropies between the excited state and the vacuum is computed by a local correlation function \cite{Alcaraz:2011tn}, which can be computed from the CFT data. To obtain the entanglement entropy, one needs to perform an analytic continuation in the \ren index $n$, which can be achieved term by term in an OPE expansion. This framework thus provides a bridge between the microscopic CFT data in a large $c$ expansion and quantum extremal surfaces in the bulk. %and we will develop a precise dictionary to go from one to the other.

At order $c^0$, this was studied in \cite{Belin:2018juv,Belin:2019mlt}. The CFT computation is given by a correlator of generalized free fields and in the bulk, one considers the entanglement entropy of free propagating bulk fields through the classical RT surface. A perfect match between bulk and boundary was found, providing an explicit check of the FLM formula \cite{Faulkner:2013ana}. It is important to emphasize that at this order, quantum extremality does not play any role since the bulk entangling surface does not move. Studying the effects that appear at order $c^{-1}$ where quantum extremality kicks in will be the focus of this paper.

The bulk side of the story turns out to be both conceptually and technically involved. Already at order $c^0$, there are contributions that appear in the area operator that exactly cancel against other contributions in the bulk entanglement entropy \cite{Belin:2018juv}. This is guaranteed by a first law of entanglement entropy in the bulk, but already suggests that there is more going on in the bulk than what the boundary theory has access to. Things get even more complicated at the next order, and we will thus start by focusing strictly on the contribution of the area operator.\footnote{Technically speaking, we do not give the full answer for the area operator as the shift of the surface due to the bulk entanglement entropy is a piece we leave for \cite{ourpapertoappear}. As we will explain, it turns out that we can separate the two effects consistently since the shift in the location of the surface picks up additive contributions from the geometry and from the bulk entanglement entropy.} The bulk entanglement entropy will be discussed in \cite{ourpapertoappear}.

We provide the entries of this dictionary in Table \ref{tab:dict}. The first two rows were derived in \cite{Belin:2018juv} while the first line of the $\mathcal{O}(c^{-1})$ dictionary is what will be derived in this paper. The anomalous dimensions $\gamma_{n,\ell}$ and corrections to mean field theory OPE coefficients of double-trace operators $a_{n,\ell}$ must involve the bulk entanglement entropy, but the precise details of this map are yet to be determined \cite{ourpapertoappear}. There could also be corrections due to the exchange of higher spin operators, giving $1/\Delta_{\mathrm{gap}}$ effects in the CFT. In the bulk, they would be responsible for deviations of the low-energy EFT from semi-classical general relativity minimally coupled to matter. Such effects would appear as corrections in any top-down model whose bulk dual is given by string theory in AdS. This part of the dictionary would be particularly interesting to construct as it would give $\alpha'$ corrections to the quantum HRRT formula \eqref{eq:HRRT}, and could probe entanglement in string theory beyond higher derivative corrections to entanglement entropy \cite{Dong:2013qoa,Camps:2013zua}. We leave it for future work.

\begin{table}[h]
\begin{center}
\begin{tabular}{c|c|c}
\multicolumn{1}{c}{}  &   \multicolumn{1}{c|}{$S_{\mathrm{EE}}^{\text{CFT}}(A)$} &   \multicolumn{1}{c}{\ext($S_{\mathrm{gen}}$)}\\
\cline{2-3}
\multirow{2}{*}{$\mathcal{O}(c^{0})$} & $\mathrm{Id}|_{h}$ & $A[\Sigma_{A}^{(0)},g^{(1)}]$ \\
    & $[OO]_{n,\ell}$ & $S^{\text{bulk}}_{\mathrm{EE}}(\Sigma_A^{(0)})$ \\ \hline
\multirow{3}{*}{$\mathcal{O}(c^{-1})$} & $\mathrm{Id}|_{\delta h} + T$ & $A[\Sigma_{A}^{(0)},g^{(2)}] + A[\Sigma_{A,\textrm{geo}}^{(1)},g^{(1)}]$ \\
    & $\gamma_{n,\ell}$ & ? \\
    & $a_{n,\ell}$ & ? \\
\hline
\multirow{3}{*}{$\mathcal{O}\left(\frac{1}{\Delta_{\mathrm{gap}}}\right)$} & & \\
    & ? & ? \\
    & & \\
\end{tabular}
\end{center}
\caption{The dictionary between CFT OPE data and the contributions to the generalized entropy of the quantum extremal surface. The terms in the $\ext(S_{\mathrm{gen}})$ column correspond to the contribution left over after the various cancellations between bulk terms.}
\label{tab:dict}
\end{table}

By matching the various contributions on the two sides of the quantum HRRT formula \eqref{eq:HRRT}, we will eventually obtain an explicit check of this formula which, to our knowledge, has not been done in the literature thus far (although see \cite{Geng:2021iyq} for an explicit check in a doubly holographic setup).

%~~~~~~~~~~~~~~~~~~~~~~~~~~~~~~~~~~~~~~~~~~~~~~~
\subsection{Summary of results}
\label{sec:Summary}
%~~~~~~~~~~~~~~~~~~~~~~~~~~~~~~~~~~~~~~~~~~~~~~

We now list a summary of results. For a scalar single-trace operator whose bulk dual is given by a free scalar field minimally coupled to gravity, the difference of CFT entanglement entropies between the excited state and the vacuum at order $c^{-1}$ is given by
\begin{equation}\label{eq:CFTresultintro}
\begin{split}
\Delta S_{\mathrm{EE}}^{\text{CFT}} &= \frac{2\delta h}{c}\left(2- \theta \cot \left(\frac{\theta}{2}\right) \right) -\frac{16h^2}{15c}\left(\sin \frac{\theta}{2}\right)^4 \\
&\qquad+ \frac{24h^2-2 \delta h }{c}\Bigg[ 2 \log \left(\sin\frac{\theta }{2}\right) \left(\sin\frac{\theta }{2}\right)^{8h} \frac{\Gamma\left(\frac{3}{2}\right)\Gamma\left(4h+1\right)}{\Gamma\left(4h+\frac{3}{2}\right)} \\
&\qquad+ \left(\sin\frac{\theta }{2}\right)^{8h} \Gamma\left(\frac{3}{2}\right) \frac{\Gamma\left(4h+1\right) }{\Gamma\left(4h+\frac{3}{2}\right)}\left(\psi(4h+1)-\psi\left(4h+\frac{3}{2}\right)\right)\Bigg] \\
&\qquad+\frac{96h^2}{c} \left(\sin\frac{\theta }{2}\right)^{8h} \frac{\Gamma\left(\frac{3}{2}\right)\Gamma\left(4h+1\right)}{\Gamma\left(4h+\frac{3}{2}\right)} \,,
\end{split}
\end{equation}
where $\delta h$ corresponds to the anomalous dimension of the single-trace operator and is not fixed from first principles in the CFT. It can be determined through the bulk and we will expand on this in the main text (see section \S\ref{sec:bulkenergy}).

The first line comes solely from the vacuum Verma module, namely the first term comes from the identity operator while the second term comes from the stress-tensor exchange. We will show that this can be reproduced in the bulk entirely from the area operator. This fits in nicely with observations for heavy states \cite{Asplund:2014coa} where it was shown that the Virasoro identity block captures the minimal area of black hole microstates or conical defects. The remaining three lines in \eqref{eq:CFTresultintro} are related to multi-trace exchanges, and the bulk entanglement entropy is needed to capture them \cite{ourpapertoappear}.

Obtaining the value of the extremal area in the bulk requires a careful treatment of semi-classical gravity coupled to matter fields, up to second order corrections in $G_N$ (which means backreacting twice). We give a precise definition of what is meant by perturbative states to this order, and take into account all effects relevant for the generalized entropy. We then compute the expectation value of the bulk area operator, reproducing the first line of \rref{eq:CFTresultintro}. In the process, many other terms appear which must cancel against contributions from the bulk entanglement entropy, as was observed to first order in \cite{Belin:2018juv}. This confirms that there is much more going on in the bulk than what the CFT has access to.

This paper is organized as follows. In section \ref{sec:CFT}, we perform the CFT side of the computation. In section \ref{sec:BulkEFT}, we discuss how to perform second-order quantum backreaction on the geometry and describe the metric that will be relevant to compute the extremal area. In section \ref{sec:Area}, we perform the extremization of the area and give the contribution of the area operator. We conclude with some open questions in section \ref{sec:discussion}. Many of the details of the calculations are provided in Appendices \ref{sec:bulkwfn+metricbackreact} and \ref{sec:areacalcs}.

%~~~~~~~~~~~~~~~~~~~~~~~~~~~~~~~~~~~~~~~~~~~~~~~
\section{CFT calculation}
\label{sec:CFT} 
%~~~~~~~~~~~~~~~~~~~~~~~~~~~~~~~~~~~~~~~~~~~~~~

In this section, we will proceed to compute the CFT entanglement entropy for perturbative excited states in a $1/c$ expansion. We start by reviewing the basics of entanglement entropy in two-dimensional conformal field theories. For more details, we refer the reader to \cite{Calabrese:2004eu,2009JPhA...42X4005C,Alcaraz:2011tn,Sarosi:2016oks,Sarosi:2017rsq,Belin:2018juv}. 

%~~~~~~~~~~~~~~~~~~~~~~~~~~~~~~~~~~~~~~~~~~~~~~~
\subsection{Entanglement entropy in CFT$_2$}
\label{sec:CFTEE}
%~~~~~~~~~~~~~~~~~~~~~~~~~~~~~~~~~~~~~~~~~~~~~~

Consider a two-dimensional conformal field theory in a state $\ket{\psi}$ with Hilbert space $\mathcal{H}$. Now imagine dividing the Hilbert space into two spatial subsystems, $A$ and its complement $\bar{A}$. To characterise the entanglement between $A$ and $\bar{A}$, we define the reduced density matrix
\begin{equation}\label{eq:densitymat}
\rho_A \equiv \Tr _{\bar{A} }\ket{\psi}\bra{\psi} \,.
\end{equation}
The entanglement entropy is given by the Von Neumann entropy of the reduced density matrix
\begin{equation}\label{eq:EE}
S_{\rm EE} = - \Tr \rho_A \log \rho_A \,.
\end{equation}
There are also other measures of entanglement, such as the \ren entropies
\begin{equation}\label{eq:Renyi}
S_n \equiv \frac{1}{1-n} \log \Tr \rho_A ^ n \,,
\end{equation}
which provide the moments of the eigenvalue distribution of $\rho_A$. Since a direct computation of the entanglement entropy is often difficult in quantum field theory, one can proceed by means of the replica trick \cite{Calabrese:2004eu,2009JPhA...42X4005C}. We compute the \ren entropies for all $n$ and then analytically continue in $n$ to obtain the entanglement entropy:\footnote{There can be subtleties in the analytic continuation \cite{2009PhRvB..80k5122M,Belin:2013dva,Belin:2017nze,Dong:2018esp,Hampapura:2018uho}, but we do not except any effect of this type in the setup relevant for this paper.}
\begin{equation}\label{eq:RenyitoEE}
S_{\rm EE} = \lim_{n\to1} S_n \,.
\end{equation}
Now consider the CFT to live on a circle of unit radius parameterized by a coordinate $\varphi$ and define the subsytem $A$ to be the spatial interval
\begin{equation}\label{eq:int}
A: \varphi \in [0,\theta] \,.
\end{equation}
We will be interested in a particular class of states $\ket{\psi}$, obtained by acting with a Virasoro primary operator on the vacuum
\begin{equation}\label{eq:ket}
\ket{\psi}=O(0)\ket{0} \,,
\end{equation}
for a primary operator $O$ with dimension $(h,\bar{h})$. The dual state is given by
\begin{equation}\label{eq:bra}
\bra{\psi}=\lim_{z\to\infty} \bra{0}O(z) z^{2h}\bar{z}^{2\bar{h}} \,.
\end{equation}
To implement the replica trick in two-dimensional CFTs, one considers the orbifold CFT $\mathcal{C}^{\otimes n}/\mathbb{Z}_n$, where the \ren entropies are given by correlation functions of twist operators \cite{Calabrese:2004eu,2009JPhA...42X4005C,Headrick:2010zt}:
\begin{equation}\label{eq:Renyi_exc}
S_n=\frac{1}{1-n} \log\bra{O^{\otimes n}}\sigma_n(0,0) \bar{\sigma}_n(0,\theta)\ket{O^{\otimes n}} \,.
\end{equation}
Note that the operator $O$ that creates the state is raised to the $n$-th power, since the replica trick instructs us to prepare $n$ copies of the state.
In quantum field theory, the entanglement and \ren entropies are UV-divergent quantities. A nice UV-finite quantity consists of the difference of entanglement entropies between the excited state and the vacuum. We have
\begin{equation}\label{eq:Reny_diff}
\Delta S_n \equiv S_n^{\text{ex}}-S_n^{\text{vac}}=\frac{1}{1-n} \log \frac{\Tr \rho_A^n}{\Tr \rho_{A,\text{vac}}^n}=\frac{1}{1-n} \log\frac{\bra{O^{\otimes n}}\sigma_n(0,0) \bar{\sigma}_n(0,\theta)\ket{O^{\otimes n}}}{\bra{0}\sigma_n(0,0) \bar{\sigma}_n(0,\theta)\ket{0}} \,.
\end{equation}
While it is possible to analyze this correlation function directly in the orbifold theory, we will proceed in a different way which will be more convenient for our purposes. We can perform a uniformization map that takes us to the covering space on the plane
\begin{equation}\label{eq:uniformcoord}
z=\left(\frac{e^w-1}{e^w-e^{i\theta}}\right)^{\frac{1}{n}} \,.
\end{equation}
In doing so, the twist operators disappear and we are left with a local correlation function in the original CFT $\mathcal{C}$ of $2n$ operators $O$ inserted on the complex plane at the positions
\begin{equation}\label{eq:kroots}
z_k=e^{-i(\theta -2\pi  k)/n}, \qquad \tilde{z}_k =e^{2\pi i  k/n} \,,  \qquad k=0, ... , n-1 \,.
\end{equation}
The difference of \ren entropies thus becomes
\begin{equation}\label{eq:diffRenyi}
\Delta S_n = \frac{1}{1-n}\log\left[e^{-  i \theta (h-\bar{h})}  \left(\frac{2}{n}\sin\left[\frac{\theta}{2}\right]\right)^{2n(h+\bar{h})} \braket{\prod_{k=0}^{n-1}O(\tilde{z}_k)O(z_k)}\right] \,.
\end{equation}
In general CFTs, the computation of this correlation function is difficult and involves all of the CFT data. For holographic CFTs with a large central charge $c$, things drastically simplify as we will now explain.

%~~~~~~~~~~~~~~~~~~~~~~~~~~~~~~~~~~~~~~~~~~~~~~~
\subsection{Holographic CFTs and the $1/c$ expansion}
\label{sec:HoloCFT}
%~~~~~~~~~~~~~~~~~~~~~~~~~~~~~~~~~~~~~~~~~~~~~~

We have seen that the \ren entropies of primary excited states are given by $2n$-point local correlation functions. As mentioned above, these are in general very hard to compute. However, we are interested in holographic large $c$ CFTs in which case the correlation function can be expanded order by order in $1/c$. To first order, the correlation function will factorize due to large $c$ factorization \cite{ElShowk:2011ag}: In large $c$ CFTs, there are two classes of operators: single-trace operators (dual to propagating bulk fields) and multi-trace operators (dual to multi-particle states in the bulk). \\
Large $c$ factorization is a property of OPE coefficients. It states that
\begin{equation}\label{eq:ST_OPE}
C_{O_1O_2O_3} \sim \frac{1}{\sqrt{c}} \,,
\end{equation}
for $O_{1,2,3}$ three single-trace operators. On the other hand, OPE coefficients involving multi-trace operators can be order one. We will consider excited states corresponding to single-particle states in the bulk and we will therefore take the primary operator $O$ to be single-trace.\\
Thanks to large $c$ factorization, the leading order contribution to the $2n$-point function \eqref{eq:diffRenyi} is given by all the Wick contractions
\begin{equation}\label{eq:GFF}
\braket{\prod_{k=0}^{n-1}O(\tilde{z}_k)O(z_k)}=\sum_{g\in S_{2n}} \prod_{j=1}^n\braket{O(z_{g(2j-1)})O(z_{g(2j)})}+\mathcal{O}\left(\frac{1}{c}\right) \,.
\end{equation}
The first correction to the entanglement entropy (which is $\mathcal{O}(1)$) reads
\begin{equation}\label{eq:EE_1stcorr}
\Delta S_n^{(1)} = \left(\frac{1}{n}\sin\left[\frac{\theta}{2}\right]\right)^{2n(h+\bar{h})} \text{Hf}(M_{ij}) \,,
\end{equation}
where $\text{Hf}(M)$ is the Haffnian of a matrix $M$ defined by
\begin{equation}\label{eq:Haff}
\text{Hf}(M)= \frac{1}{2^n n!} \sum_{g\in S_{2n}}\prod_{j=1}^n M_{g(2j-1),g(2j)} \,,
\end{equation}
and
\begin{equation}\label{eq:Mmatrix}
M_{ij}=\begin{cases}
\frac{1}{(| \sin \frac{\pi (i-j)}{n}|)^{2(h+\bar{h})}}, \qquad i,j\leq n \\
\frac{1}{\left( |\sin \left( \frac{\pi (i-j)}{n}-\frac{\varphi}{2n}\right)|\right)^{2(h+\bar{h})}}, \qquad i\leq n, \quad j>n \\
\frac{1}{\left( |\sin \left( \frac{\pi (i-j)}{n}+\frac{\varphi}{2n}\right)|\right)^{2(h+\bar{h})}}, \qquad j\leq n, \quad i>n \\
\frac{1}{( |\sin \frac{\pi (i-j)}{n}|)^{2(h+\bar{h})}}, \qquad i,j> n \,.
\end{cases}
\end{equation}
This is the exact expression for the \ren entropies in primary single-trace states to order $c^0$. Unfortunately, it is difficult to perform the analytic continuation $n\to1$ for general $h$ and $\bar{h}$ (there is however an analytic continuation for $h+\bar{h}=1$ \cite{Berganza:2011mh,PhysRevLett.110.115701,1742-5468-2014-9-P09025,Ruggiero:2016khg}). To simplify the analytic continuation, we will consider a small interval limit where one performs an OPE expansion in the CFT. From now on we consider scalar states and set $h=\bar{h}$.

%~~~~~~~~~~~~~~~~~~~~~~~~~~~~~~~~~~~~~~~~~~~~~~~
\subsection{Small interval limit}
\label{sec:smallint}
%~~~~~~~~~~~~~~~~~~~~~~~~~~~~~~~~~~~~~~~~~~~~~~

The small interval ($\theta \ll 1$) limit  corresponds to the OPE limit in the $2n$-point correlation function, where the $2n$ operators come close pairwise. The leading term corresponds to the identity contribution for all OPE contractions and reads
\begin{equation}\label{eq:Renyi_identcont}
\Delta S_n \approx \left(\frac{\sin\left[\frac{\theta}{2}\right]}{n\sin\left[\frac{\theta}{2n}\right]}\right)^{4nh} \,.
\end{equation}
The analytic continuation for this term can easily be computed and gives
\begin{equation}\label{eq:EE_identcont} 
\Delta S_{\rm EE} \approx 2h\left(2- \theta \cot \left(\frac{\theta}{2}\right) \right) \,.
\end{equation}
The right-hand side of this expression is actually fixed by conformal invariance, and is exactly equal to the expectation value of the vacuum modular Hamiltonian in the primary excited state (see for example \cite{Sarosi:2016oks}). Beyond this order, we probe the generalized free fields as explained in the previous subsection. From the OPE point of view, all the other Wick contractions  come from the contribution of multi-trace operators.
The lightest operator that appears in the $O \times O$ OPE is the operator $:O^2:$ with conformal dimension $4h$. Its OPE coefficient is given by \cite{Belin:2017nze}
\begin{equation}\label{eq:O^2OPEcoeff}
C_{OO O^2}=\sqrt{2} \,.
\end{equation}
The exchange of $O^2$ by any two pairs of the $2n$-point function gives the subleading contribution\footnote{Strictly speaking, the $\theta^{8h}$ term is subleading to the $\theta\cot \left(\frac{\theta}{2}\right)$ term only for $h>\frac{1}{4}$. However, it is essential for our analysis to organize terms in the small interval limit by their analytic behavior in the complex $\theta$-plane and keep all terms with different analytic behavior.}
\begin{equation} \label{eq:CFTanswer_1storder}
\Delta S_{\mathrm{EE}}|_{\mathcal{O}(c^{0})} = 2h\left(2- \theta \cot \left(\frac{\theta}{2}\right) \right) - \left(\sin\frac{\theta }{2}\right)^{8h} \frac{\Gamma\left(\frac{3}{2}\right)\Gamma\left(4h+1\right)}{\Gamma\left(4h+\frac{3}{2}\right)}\ + \cdots.
\end{equation}
This result was reproduced by a bulk calculation in \cite{Belin:2018juv}. We now turn to higher order corrections in the $1/c$ expansion.

\subsection{$1/c$ corrections}

We are now ready to discuss the further $1/c$ corrections. These $1/c$ corrections can come from four different sources:
\begin{enumerate}
\item The operator $O$ that we consider may not have a protected conformal dimension, and it can receive $1/c$ corrections. We will therefore write\footnote{With a slight abuse of notation, we will use $h$ as the bare (infinite $c$) scaling dimension and not as $\Delta^{\textrm{full}}/2$.}
\begin{equation}\label{eq:Odim}
\left(h^{\textrm{full}},\bar{h}^{\textrm{full}}\right)_O= \left(h + \frac{\delta h}{c},h + \frac{\delta h}{c}\right) \,.
\end{equation}

\item The exchange of other single-trace operators. In particular, an operator that always contributes is the exchange of the stress-tensor. In this work, we will assume no other operator appears in the OPE of $O$ with $O$. In particular, this means we are considering a theory where $C_{OOO}=0$ (for example because $O$ is protected by a $\mathbb{Z}_2$ symmetry).
\item Anomalous dimensions of the multi-trace operators:
\begin{equation}\label{eq:DT_anomdim}
\Delta_{[OO]_{n,l}}=4h + 2n + l +\frac{\gamma_{n,l}+4  \delta h}{c} +\mathcal{O}(c^{-2})
\end{equation}
These anomalous dimensions can come from two sources. They can either come from truncated solutions to crossing, which can be thought of as bulk quartic couplings \cite {Heemskerk:2009pn}, or they can be induced by crossing due to the exchange of single-trace operators \cite{Alday:2017gde}. We will assume the bulk scalar sector is free and thus does not have quartic couplings, so the sole source of anomalous dimensions is coming from the crossing data due to the exchange of the stress tensor $T$ (and in principle all Virasoro descendants of the identity, although the stress-tensor is the only contribution at this order).
 
\item Changes in the OPE coefficients of the multi-trace operators:
\begin{equation}\label{eq:DT_OPEcoeff}
C_{OO[OO]_{n,l}}^2=(1+(-1)^l)C_nC_{n+l}+\frac{a_{n,l}}{c} +\mathcal{O}(c^{-2}) \,, 
\end{equation}
with $C_n=\frac{\Gamma(2h+n)^2\Gamma(4h+n-1)}{n!\Gamma(2h)^2\Gamma(4h+2n-1)}$.
As with the anomalous dimensions, these can come from two sources and we will assume they are only generated by the crossing data of the stress-tensor exchange.
\end{enumerate}

We will now discuss the effect of each contribution. We start with the anomalous dimension of the single-trace operator. This contribution is easily obtained from the result at order $c^0$ by shifting $h\to h+ \delta h/c$. Keeping only the contribution of the identity operator since the double-trace operator will be dealt with shortly, we find
\begin{equation} \label{eq:CFTanswer_STanom}
\Delta S_{\mathrm{EE}}^1 = \frac{2\delta h}{c}\left(2- \theta \cot \left(\frac{\theta}{2}\right) \right) \,.
\end{equation}
Next, we consider the exchange of the stress tensor, and proceed in a similar manner to how we obtained \eqref{eq:CFTanswer_1storder}. The (unnormalized) OPE coefficient is given by
\begin{equation}\label{eq:TOPEcoeff}
C_{OOT}=h \,,
\end{equation}
So the exchange of $T$ in the $2n$-point function gives a contribution
\begin{equation}\label{eq:Tcontr}
\frac{2h^2}{c}\left(\frac{\sin\left[\frac{\theta}{2}\right]}{n\sin\left[\frac{\theta}{2n}\right]}\right)^{4nh} \left(\left(\sin\frac{\theta }{2n}\right)^{4}\ \sum_{k=1}^{n-1}\frac{n-k}{(\sin\frac{\pi k }{n})^{4}}  \right)   \,.
\end{equation}
Using the periodicity of the sine function, this can be rewritten as
\begin{equation}\label{eq:Tcontrrewrite}
\frac{h^2}{c}\left(\frac{\sin\left[\frac{\theta}{2}\right]}{n\sin\left[\frac{\theta}{2n}\right]}\right)^{4nh} \left(\left(\sin\frac{\theta }{2n}\right)^{4}\ n \sum_{k=1}^{n-1}\frac{1}{(\sin\frac{\pi k }{n})^{4}}  \right)   \,.
\end{equation}
The analytic continuation of this expression was obtained in \cite{Calabrese:2010he} and reads
\begin{equation}\label{eq:Tcontr_analcont}
n\sum_{k=1}^{n-1}\frac{1}{(\sin\frac{\pi k }{n})^{4}} =(n-1)\frac{\Gamma\left(\frac{3}{2}\right)\Gamma\left(3\right)}{\Gamma\left(\frac{7}{2}\right)}+\mathcal{O}((n-1)^2)=(n-1)\frac{8}{15}+\mathcal{O}((n-1)^2) \,.
\end{equation}
We can then obtain the contribution of $T$ and $\bar{T}$ to the entanglement entropy
\begin{equation}\label{eq:CFTanswer_T}
\Delta S_{\mathrm{EE}}^2=-\frac{16h^2}{15c}\left(\sin \frac{\theta}{2}\right)^4 \,.
\end{equation}
Note that this is also the subleading contribution to the identity \textit{Virasoro} block, which matches the computation done in \cite{Asplund:2014coa}. 

The third piece comes from the shift in the dimension of $O^2$. We can easily find this contribution to be
\begin{equation}\label{eq:CFTanswer_DTanom}
\begin{split}
\Delta S_{\mathrm{EE}}^3 &= - \frac{\gamma_{0,0}+4 \delta h }{c}\Bigg[ 2 \log \left(\sin\frac{\theta }{2}\right) \left(\sin\frac{\theta }{2}\right)^{8h} \frac{\Gamma\left(\frac{3}{2}\right)\Gamma\left(4h+1\right)}{\Gamma\left(4h+\frac{3}{2}\right)} \\
	&+ \left(\sin\frac{\theta }{2}\right)^{8h} \Gamma\left(\frac{3}{2}\right) \frac{\Gamma\left(4h+1\right) }{\Gamma\left(4h+\frac{3}{2}\right)}\left(\psi(4h+1)-\psi\left(4h+\frac{3}{2}\right)\right)\Bigg]
\end{split}
\end{equation}
where $\psi$ is the digamma function.

Finally, the contribution due to the change in OPE coefficients reads
\begin{equation}\label{eq:CFTanswer_DTOPE}
\Delta S_{\mathrm{EE}}^4=-\frac{a_{0,0}}{c} \left(\sin\frac{\theta }{2}\right)^{8h} \frac{\Gamma\left(\frac{3}{2}\right)\Gamma\left(4h+1\right)}{\Gamma\left(4h+\frac{3}{2}\right)} \,.
\end{equation}

It remains to give the values of $\gamma_{n,l}$ and $a_{n,l}$. These were computed in \cite{Collier:2018exn}. We have
\begin{equation}\label{eq:DTanomdim}
\gamma_{0,0}=-24h^2\,.
\end{equation}
The value $a_{0,0}$ can be extracted from (B.1) of \cite{Collier:2018exn}, we find
\begin{equation}\label{eq:DTOPE}
C_{OO[OO]_{0,0}}^2=2\frac{\Gamma_b(2Q)}{\Gamma_b(Q)} \frac{\Gamma_b(2Q-4\alpha)\Gamma_b(Q-2\alpha)^2}{\Gamma_b(Q-4\alpha)\Gamma_b(2Q-2\alpha)^2}  \,,
\end{equation}
where $\Gamma_b$ is the Barnes double Gamma function and
\begin{equation}\label{eq:Lioudata}
\begin{split}
c &= 1+6 Q^2 \\
Q &= b+b^{-1} \\
h &= \alpha(Q-\alpha)
\end{split}
\end{equation}
Expanding this at large $c$ and fixed $h$, we find 
\begin{equation}\label{eq:DTOPEcoeff}
a_{0,0}=-96h^2 \,.
\end{equation}
Putting all the pieces together, we find that at order $1/c$, the difference of entanglement entropies is given by
\begin{equation}\label{eq:CFTresult_2ndorder}
\begin{split}
\Delta S_{\mathrm{EE}}|_{\mathcal{O}(c^{-1})} &= \frac{2\delta h}{c}\left(2- \theta \cot \left(\frac{\theta}{2}\right) \right) -\frac{16h^2}{15c}\left(\sin \frac{\theta}{2}\right)^4 \\
	&+\frac{24h^2-4 \delta h }{c}\Bigg[ 2 \log \left(\sin\frac{\theta }{2}\right) \left(\sin\frac{\theta }{2}\right)^{8h} \frac{\Gamma\left(\frac{3}{2}\right)\Gamma\left(4h+1\right)}{\Gamma\left(4h+\frac{3}{2}\right)} \\
	&+ \left(\sin\frac{\theta }{2}\right)^{8h} \Gamma\left(\frac{3}{2}\right) \frac{\Gamma\left(4h+1\right) }{\Gamma\left(4h+\frac{3}{2}\right)}\left(\psi(4h+1)-\psi\left(4h+\frac{3}{2}\right)\right)\Bigg] \\
	&+\frac{96h^2}{c} \left(\sin\frac{\theta }{2}\right)^{8h} \frac{\Gamma\left(\frac{3}{2}\right)\Gamma\left(4h+1\right)}{\Gamma\left(4h+\frac{3}{2}\right)} \,.
\end{split}
\end{equation}
The aim will now be to reproduce this result from the bulk side, and understand how the various pieces whose origin is clear in the CFT get reorganized and geometrized in the bulk theory. In this paper, we will concentrate on reproducing the first line of \eqref{eq:CFTresult_2ndorder}, which we will show is solely encoded in the area operator. The bulk entanglement entropy responsible for the other terms will be discussed in \cite{ourpapertoappear}.

Before moving on, two comments are in order. First, this expression is meant in a double-expansion sense. It is the expansion of the full answer to order $1/c$, further expanded in the small interval limit where only the leading term in $\theta$ is kept for each piece. Second, it is striking to see that the anomalous dimension of the single-trace operator is not fixed by crossing in any way (at least in the perturbative sense), and depends on details of the microscopic theory at hand. This is similar to bulk quartic couplings that may or may not be there, depending on the bulk theory. We will see that this has a direct counterpart in the bulk, and that the CFT operator can be protected or not.

%~~~~~~~~~~~~~~~~~~~~~~~~~~~~~~~~~~~~~~~~~~~~~~~
\section{Bulk effective field theory}
\label{sec:BulkEFT} 
%~~~~~~~~~~~~~~~~~~~~~~~~~~~~~~~~~~~~~~~~~~~~~~

Our ultimate goal is to compute the expectation value of the area operator in the state $\ket{\psi}_{\mathrm{bulk}}$ evaluated on the quantum extremal surface up to $\mathcal{O}(G_{N}^{2})$. This requires backreacting the corresponding matter on the spacetime twice to obtain $\mathcal{O}(G_{N}^{2})$ corrections to the metric, which will be the subject of this section.  We will consider pure Einstein gravity minimally coupled to a scalar field.

%~~~~~~~~~~~~~~~~~~~~~~~~~~~~~~~~~~~~~~~~~~~~~~~
\subsection{First-order backreaction}
\label{sec:1storderbackreact}
%~~~~~~~~~~~~~~~~~~~~~~~~~~~~~~~~~~~~~~~~~~~~~~

The first order backreaction of the bulk state $\ket{\psi}_{\mathrm{bulk}}$ on the metric was obtained in \cite{Belin:2018juv} which we review here. We start with pure AdS$_{3}$ in global coordinates
\begin{equation}\label{eq:AdS3metric}
(ds^{(0)})^{2} = -(r^{2}+1)dt^2+\frac{dr^2}{r^{2}+1}+r^2d\varphi^2 \,.
\end{equation}
Consider a free massive scalar field $\phi^{(0)}$ propagating on this background. The wave equation for this scalar field is
\begin{equation}\label{eq:waveeqn_0thorder}
\left((\nabla^{(0)})^{2}-m^2\right)\phi^{(0)} = 0 \,,
\end{equation}
where the mass $m$ is related to the conformal dimension $h$ of the CFT operator $O$ dual to $\phi$ via
\begin{equation}\label{eq:mass-dimension}
m^2 = 4h(h-1) \,.
\end{equation}
We canonically quantize the field
\begin{equation}\label{eq:canonicalquant_0thorder}
\phi^{(0)}(t,r,\varphi) = \sum_{n,m}\left(a_{n,m}^{(0)}e^{-i\Omega_{n,m}^{(0)}t}f_{n,m}^{(0)}(r,\varphi)+\left(a_{n,m}^{(0)}\right)^{\dagger}e^{i\Omega_{n,m}^{(0)}t}\left(f_{n,m}^{(0)}\right)^{\ast}(r,\varphi)\right) \,,
\end{equation}
where $e^{-i\Omega_{n,m}t}f_{n,m}$ satisfy the wave equation \eqref{eq:waveeqn_0thorder} and the ladder operators obey the canonical commutation relations
\begin{equation}\label{eq:ladderops_0thorder}
[a_{n,m}^{(0)},\left(a_{n',m'}^{(0)}\right)^{\dagger}] = \delta_{mm'}\delta_{nn'} \,.
\end{equation}
We are interested in the state with the lowest energy excitation $(n=m=0)$ of the scalar field $\phi$, namely
\begin{equation}\label{eq:state} 
\ket{\psi^{(0)}}_{\mathrm{bulk}} = \left(a_{0,0}^{(0)}\right)^{\dagger}\ket{0^{(0)}} \,,
\end{equation}
which corresponds to a primary operator in the CFT.\footnote{We will henceforth refer to $\ket{\psi}_{\mathrm{bulk}}$ simply as $\ket{\psi}$.} To solve the wave equation \eqref{eq:waveeqn_0thorder}, we expand the wavefunction $f_{n,m}^{(0)}(r,\varphi)$ in a Fourier series
\begin{equation}\label{eq:Fourierexp}
f_{n,m}^{(0)}(r,\varphi) = \sum_{m}e^{2\pi im\varphi}f_{n,m}^{(0)}(r) \,.
\end{equation}
The solution to the wave equation for $n=m=0$ after imposing regularity at the origin, normalizability at the boundary, and unit norm is given by
\begin{equation}\label{eq:lowestwavefn_0thorder}
f_{0,0}^{(0)}(r) = \frac{1}{\sqrt{2\pi}}\frac{1}{(1+r^2)^{h}}, \qquad \Omega_{0,0}^{(0)} = 2h \,.
\end{equation}
To determine the back-reaction of this excitation, we need the expectation value of the stress-tensor in the state $\ket{\psi^{(0)}}$. The stress-tensor is given by
\begin{equation}\label{eq:stress-tensor}
T_{\mu\nu}^{(0)} = \;\normord{\partial_{\mu}\phi^{(0)}\partial_{\nu}\phi^{(0)}-\frac{1}{2}g_{\mu\nu}^{(0)}\left((\nabla^{(0)}\phi^{(0)})^2+m^2(\phi^{(0)})^2\right)} \,,
\end{equation}
whose expectation values are
\begin{equation}\label{eq:stress-tensor_evs_0thorder}
\begin{split}
\bra{\psi^{(0)}}T_{tt}^{(0)}\ket{\psi^{(0)}} &= \frac{2h(2h-1)}{\pi}\frac{1}{(1+r^2)^{2h-1}} \\
\bra{\psi^{(0)}}T_{rr}^{(0)}\ket{\psi^{(0)}} &= \frac{2h}{\pi}\frac{1}{(1+r^2)^{2h+1}} \\
\bra{\psi^{(0)}}T_{\varphi\varphi}^{(0)}\ket{\psi^{(0)}} &= \frac{2hr^2}{\pi}\frac{(1-2h)r^2+1}{(1+r^2)^{2h+1}} \,.
\end{split}
\end{equation}
The first-order backreacted metric is then obtained from Einstein's equations:
\begin{equation}\label{eq:EE_0thorder}
\left(R_{\mu\nu} - \frac{1}{2}g_{\mu\nu}R-g_{\mu\nu}\right)\bigg|_{\mathcal{O}(G_{N})} = 8\pi G_{N}\bra{\psi^{(0)}}T_{\mu\nu}^{(0)}\ket{\psi^{(0)}} \,,
\end{equation}
where we expand the metric in powers of $G_{N}$
\begin{equation}\label{eq:metricexp}
g_{\mu\nu} = g_{\mu\nu}^{(0)} + G_{N}g_{\mu\nu}^{(1)} + G_{N}^{2}g_{\mu\nu}^{(2)} + \ldots
\end{equation}
with $g_{\mu\nu}^{(0)}$ given by \eqref{eq:AdS3metric}, and we have similar expansions for $R_{\mu\nu}$ and $R$. The solution of Einstein's equations up to $\mathcal{O}(G_{N})$, after requiring that the metric be asymptotically AdS$_{3}$ and smooth at the origin, is given by
\begin{equation}\label{eq:metric_1storder}
ds^2 = -(r^2+1-16G_{N}h)dt^2+\left[1+\frac{16G_{N}h}{r^2+1}\left(1-\frac{1}{(r^2+1)^{2h-1}}\right)\right]\frac{dr^2}{r^2+1}+r^2d\varphi^2+\mathcal{O}(G_{N}^{2}) \,.
\end{equation}

We would like to emphasize that this geometry is smooth everywhere, and can be viewed as a regularized conical defect geometry, where the Compton wavelength of the particle smoothens out the conical defect.
%

%~~~~~~~~~~~~~~~~~~~~~~~~~~~~~~~~~~~~~~~~~~~~~~~
\subsection{Second-order backreaction}
\label{sec:2ndorderbackreact}
%~~~~~~~~~~~~~~~~~~~~~~~~~~~~~~~~~~~~~~~~~~~~~~

We now want to backreact the matter on the spacetime a second time. This requires quantizing a free scalar field on the first-order spacetime and backreacting yet again the excited state on top of this geometry to obtain a second-order backreacted metric.

%~~~~~~~~~~~~~~~~~~~~~~~~~~~~~~~~~~~~~~~~~~~~~~~
\subsubsection{Scalar field}
\label{sec:2ndorderbackreact_scalar}
%~~~~~~~~~~~~~~~~~~~~~~~~~~~~~~~~~~~~~~~~~~~~~~

Consider a free massive scalar field $\phi$ propagating on the first-order backreacted spacetime \eqref{eq:metric_1storder}. The wave equation is
\begin{equation}\label{eq:waveeqn_1storder}
\big(\nabla^2-m^2\big)\phi = 0 \,,
\end{equation}
where the differential operator can be written as
\begin{equation}\label{eq:nabla_def}
\begin{split}
\nabla^{2} &= \left(\nabla^{(0)}\right)^2+16hG_{N}D^{(1)}+\mathcal{O}(G_{N}^{2})
\end{split}
\end{equation}
with the differential operators $\left(\nabla^{(0)}\right)^2$ and $D^{(1)}$ provided in \eqref{eq:Laplaceop_defs}. Once again we consider the mode expansion of the wavefunction
\begin{equation}\label{eq:canonicalquant_1storder}
\phi(t,r,\varphi) = \sum_{n,m}\big(a_{n,m}e^{-i\Omega_{n,m}t}f_{n,m}(r,\varphi)+a_{n,m}^{\dagger}e^{i\Omega_{n,m}t}\big(f_{n,m}\big)^{\ast}(r,\varphi)\big)
\end{equation}
with components
\begin{equation}\label{eq:Fourierexp_1storder}
f_{n,m}(r,\varphi) = \sum_{m}e^{2\pi im\varphi}f_{n,m}(r)
\end{equation}
and canonical commutation relations
\begin{equation}\label{eq:ladderops_1storder}
[a_{n,m},a_{n',m'}^{\dagger}] = \delta_{mm'}\delta_{nn'} \,.
\end{equation}

The radial components $f_{n,m}(r)$ and frequencies $\Omega_{n,m}$ can be expanded around the pure AdS$_{3}$ solution:
\begin{equation}\label{eq:radialcompfreq_1storder}
f_{n,m}(r) = f_{n,m}^{(0)}(r) + 16hG_{N}f_{n,m}^{(1)}(r) + \mathcal{O}(G_{N}^{2}), \qquad \Omega_{n,m} = \Omega_{n,m}^{(0)}+16hG_{N}\Omega_{n,m}^{(1)} + \mathcal{O}(G_{N}^{2}) \,.
\end{equation}
The wave equation then becomes an inhomogeneous second order differential equation given by the pure AdS$_{3}$ wave equation with a source term:
\begin{equation}\label{eq:inhomoeq}
\left(\left(\nabla^{(0)}\right)^2-m^{2}\right)e^{-i\Omega_{n,m}^{(0)}t}f_{n,m}^{(1)}(r,\varphi) = -\left(D^{(1)}+\frac{2}{r^{2}+1}\Omega_{n,m}^{(0)}\Omega_{n,m}^{(1)}\right)e^{-i\Omega_{n,m}^{(0)}t}f_{n,m}^{(0)}(r,\varphi) \,.
\end{equation}
This differential equation turns out to be solvable analytically for the lowest energy wavefunction $n=m=0$. We refer the reader to App. \ref{sec:bulkwfn} for the details and here we simply state the result:
\begin{equation}\label{eq:wavefn_1storder}
\begin{split}
f_{0,0}^{(1)}(y) &= (1-y)^{h}\Bigg(\mathcal{C}_{1}+\mathcal{C}_{2}\bigg(\ln y + 2hy \,{}_{3}{F}_{2}\big(1,1,2h+1;2,2;y\big)\bigg) \\
	&-\frac{h}{2\sqrt{2\pi}}\bigg[\frac{1}{2h}(1-y)^{2h}-\frac{(2h-1)}{(4h-1)}y \,{}_{3}{F}_{2}\big(1,1,2-2h;2,2;y\big)+2y-\bigg(1+\frac{\Omega_{0,0}^{(1)}}{h}-\frac{1}{(4h-1)}\bigg)\ln y\bigg]\Bigg) \\
	y &\equiv \frac{r^{2}}{r^{2}+1} \,.
\end{split}
\end{equation}
The requirement of regularity at the origin fixes $\mathcal{C}_{2}$ and normalizability of the wavefunction gives
\begin{equation}\label{eq:freq_1storder}
\Omega_{0,0}^{(1)} = -2h\frac{(2h-1)}{(4h-1)} \,,
\end{equation}
and the requirement that the wavefunction have unit norm fixes $\mathcal{C}_{1}$. These lead to the final form for the wavefunction
\begin{equation}\label{eq:wavefn_1storder_final}
f_{0,0}^{(1)}(y) = \frac{f_{0,0}^{(0)}(y)}{2}\left(1+\frac{h}{(4h-1)}\big(\psi(2h)-\psi(4h)\big)-\frac{1}{2}(1-y)^{2h}-2hy+h\frac{(2h-1)}{(4h-1)}y\,{}_{3}{F}_{2}\big(1,1,2-2h;2,2;y\big)\right) \,.
\end{equation}
%

%~~~~~~~~~~~~~~~~~~~~~~~~~~~~~~~~~~~~~~~~~~~~~~~
\subsubsection{Metric}
\label{sec:2ndorderbackreact_metric}
%~~~~~~~~~~~~~~~~~~~~~~~~~~~~~~~~~~~~~~~~~~~~~~

We would now like to backreact this scalar field on top of the geometry \eqref{eq:metric_1storder}. The bulk state dual to $\ket{\psi}_{\mathrm{CFT}}$ is\footnote{It is worth mentioning that at order $G_N^2$ (or $c^{-1}$ in the CFT), the bulk scalar sector is no longer free due to the interactions with gravity. In particular, the energy of the two-particle state is not twice that of the one-particle state and the Fock nature of the Hilbert space disappears. Moreover, the CFT operators dual to one-particle states can mix with multi-trace operators \cite{Aprile:2020uxk,Rawash:2021pik}. However, one can check that for one-particle states, these features do not affect the area operator (the technology of \cite{Fitzpatrick:2011hh} is useful to that end). They could however affect the bulk entanglement entropy.}
\begin{equation}\label{eq:bulkstate}
\ket{\psi} = a_{0,0}^{\dagger}\ket{0} \,.
\end{equation}
We are ultimately interested in computing the area which is only sensitive to $g_{rr}^{(2)}$. This comes from the $tt$ Einstein's equation so we focus here only on solving this equation (the details of solving the other equations can be found in \S\ref{sec:metric_2ndorder}). The expectation value of the $tt$ component of the stress-tensor in this state to $\mathcal{O}(G_{N})$ is
\begin{equation}\label{eq:Ttt_1storder}
\begin{split}
\bra{\psi}T_{tt}\ket{\psi}|_{\mathcal{O}(G_{N})} &= 16hG_{N}\Bigg[4h\left(\Omega_{0,0}^{(1)}-h+1\right)\left(f_{0,0}^{(0)}\right)^2+8h\left(h+(h-1)(r^2+1)\right)f_{0,0}^{(0)}f_{0,0}^{(1)} \\
	&-\left(2(r^2+1)-\frac{1}{(r^2+1)^{2h-2}}\right)\left(\partial_{r}f_{0,0}^{(0)}\right)^{2}+2(r^2+1)^2\partial_{r}f_{0,0}^{(0)}\partial_{r}f_{0,0}^{(1)}\Bigg] \,.
\end{split}
\end{equation}
The $tt$ Einstein's equation at $\mathcal{O}(G_{N}^{2})$ is given by
\begin{equation}\label{EEtt_2ndorder}
\left(R_{tt} - \frac{1}{2}g_{tt}R-g_{tt}\right)\bigg|_{\mathcal{O}(G_{N}^{2})} = 8\pi G_{N}\bra{\psi}T_{tt}\ket{\psi}|_{\mathcal{O}(G_{N})} \,.
\end{equation}
This can be solved analytically for $g_{rr}^{(2)}$, which is explained in detail in App. \ref{sec:metric_2ndorder}, and one finds
\begin{equation}\label{eq:grr_2ndorder}
g_{rr}^{(2)} = {-}64h^{2}\frac{(r^{2}-3)}{(r^{2}+1)^{3}}\left(1-\frac{1}{(r^{2}+1)^{2h-1}}\right)^{2}-\frac{2}{(r^{2}+1)^{2}}G_{2}^{(2)}(r) \,,
\end{equation}
where
\begin{equation}\label{eq:G22def}
\begin{split}
G_{2}^{(2)}(r) &= \frac{32h^{2}}{(4h-1)}\Bigg[\frac{2}{(r^{2}+1)^{2h}}\left((4h-1)(3-4h)r^{2}+4h+2h(r^{2}+1)\left(\psi(2h)-\psi(4h)\right)\right) \\
	&-\frac{(4h-1)r^{2}+8h-3}{(r^{2}+1)^{4h-1}}+4h(2h-1)\frac{r^{4}}{(r^{2}+1)^{2h+1}}{}_{3}{F}_{2}\bigg(1,1,2-2h;2,2;\frac{r^{2}}{r^{2}+1}\bigg) \\
	&+8h^{2}\sum_{k=1}^{\infty}\frac{(1-2h)_{k}}{kk!}B_{y}(k+1,2h)\Bigg]+\mathcal{C}_{3} \,,
\end{split}
\end{equation}
where $B_{y}(a,b)$ is the incomplete Beta function. The constant of integration $\mathcal{C}_{3}$ is fixed by the requirement that there is no conical singularity at $r=0$:
\begin{equation}\label{eq:C3def}
\mathcal{C}_{3} = -\frac{32h^{2}}{(4h-1)}\left(4h\left(\psi(2h)-\psi(4h)\right)+3\right) \,.
\end{equation}
This completes the second-order backreaction of the scalar field on the spacetime that will be needed to determine the area of the quantum extremal surface coming from second order changes in the metric. Again, this geometry should be viewed as a smoothened-out conical defect, where the smoothing is due to the quantum nature of the particle we are inserting. The metric shares certain similarities with a conical defect, in particular near the boundary. We will see this directly in the value of the area.

%~~~~~~~~~~~~~~~~~~~~~~~~~~~~~~~~~~~~~~~~~~~~~~~
\subsection{Bulk energy}
\label{sec:bulkenergy}
%~~~~~~~~~~~~~~~~~~~~~~~~~~~~~~~~~~~~~~~~~~~~~~

Before moving to the evaluation of the area operator, we first need to discuss the energy of the one-particle state. As we will see, this will encode the anomalous dimension of the single-trace operator, which was not fixed from CFT first principles in \S\ref{sec:CFT} (although it would of course be fixed in any given top-down model, and is allowed to be non-zero if the operator is not protected).

At order $c^{0}$, there are two ways to think about the bulk dual of the CFT state $\ket{\psi}_{\mathrm{CFT}}$. Either we think of it as a state of the perturbative matter $\ket{\psi^{(0)}}_{\mathrm{bulk}}$ on a frozen AdS$_{3}$ background, or we can view it as dual to the geometry \eqref{eq:metric_1storder} coming from backreacting $\ket{\psi^{(0)}}_{\mathrm{bulk}}$ on top of pure AdS$_{3}$. In fact, one really needs to view the state in both ways \textit{simultaneously} to compute the generalized entropy, since the matter state affects the entanglement entropy while the backreaction affects the area. However, when trying to extract the total energy of the state, we can think of it in either way. We can compute the ADM mass using the geometry  \eqref{eq:metric_1storder}, or we can compute the bulk matter energy. We find
\be
M_{\textrm{ADM}}-M_{\textrm{AdS}}=\bra{\psi^{(0)}}H_{\mathrm{matter}}\ket{\psi^{(0)}}=2h=\Delta_O^{\textrm{CFT}} \,,
\ee
namely both procedures agree with the scaling dimension of the CFT operator, as expected. The agreement of the two procedures is guaranteed by Einstein's equations for linearized perturbation on top of AdS, but this breaks down at second order ($\mathcal{O}(c^{-1})$) as we will now see. Of course, the AdS/CFT dictionary instructs us to take the ADM mass to be dual to the CFT energy, but we will compute both quantities nevertheless.

%Therefore, as we will see momentarily, the expectation value of the scalar field Hamiltonian in the state $\ket{\psi^{(0)}}_{\mathrm{bulk}}$ and the ADM mass of the backreacted spacetime \eqref{eq:metric_1storder} both give the order $c^{0}$ conformal dimension of $O$ equal to $2h$. However, this picture breaks down at $\mathcal{O}(c^{-1})$, where only the second-order backreacted spacetime can be trusted to be dual to the CFT state $\ket{\psi}_{\mathrm{CFT}}$. In particular, we will compute the ADM mass of the second-order backreacted spacetime to extract the $\delta h/c$ correction to the conformal dimension of $O$, which we were not able to compute directly from the CFT 

The Hamiltonian of the bulk scalar field $\phi$ is related in the standard way to the $tt$ component of the bulk stress-tensor:
\begin{equation}\label{eq:Ham}
H = -\int d^{2}x\,\sqrt{-g}\,g_{tt}T^{tt} \,.
\end{equation}
From \eqref{eq:stress-tensor_evs_0thorder} and \eqref{eq:Ttt_1storder}, one finds the expectation value of the Hamiltonian to $\mathcal{O}(G_{N})$ to be
\begin{equation}\label{eq:expH}
\bra{\psi}H\ket{\psi} = \Omega_{0,0} = 2h - 48h^{2}\frac{(2h-1)}{(4h-1)}\frac{1}{c} \,,
\end{equation}
where we have used the AdS/CFT relation $c=\frac{3}{2G_{N}}$. This is the expected relation between the energy and frequency of a free scalar field. We see that the leading order term is equal to the bare conformal dimension of $O$, but there is a non-trivial correction at subleading order.

The ADM mass of any asymptotically AdS spacetime can be computed from the quasilocal stress-tensor on the timelike boundary which is defined by the variation of the gravitational action with respect to the boundary metric with suitable counter-terms added to remove the standard divergences near the boundary of AdS \cite{Balasubramanian:1999re}. This quasilocal stress-tensor is interpreted as the expectation value of the CFT stress-tensor in the CFT state dual to the spacetime geometry. Explicitly, it is given in AdS$_{3}$ by
\begin{equation}\label{eq:CFTstresstensor}
\bra{\psi}T_{\mu\nu}^{\mathrm{CFT}}\ket{\psi} = \frac{1}{8\pi G_{N}}\Big(\mathcal{K}\gamma_{\mu\nu}-\mathcal{K}_{\mu\nu}-\gamma_{\mu\nu}\Big) \,,
\end{equation}
where $\mathcal{K}_{\mu\nu}$ is the extrinsic curvature of the codimension-1 timelike surface defined by fixed $r$, which is eventually taken to infinity, and $\gamma_{\mu\nu}$ is the boundary metric. The extrinsic curvature after taking the limit $r \to \infty$ is
\begin{equation}\label{eq:extcurv}
\begin{split}
\mathcal{K}_{tt} &= -r^{2}-\frac{1}{2}+8hG_{N}-256h^{2}G_{N}^{2}\left(\frac{(2h-1)}{4(4h-1)}+\frac{h}{(4h-1)}\left(\psi(4h)-\psi(2h+1)\right)\right) \\
\mathcal{K}_{\varphi\varphi} &= r^{2}+\frac{1}{2}-8hG_{N}+64h^{2}G_{N}^{2}\frac{(2h-1)}{(4h-1)} \\
\mathcal{K} &= 2 \,,
\end{split}
\end{equation}
leading to the quasilocal stress tensor
\begin{equation}\label{eq:CFTstresstensor_final}
\bra{\psi}T_{tt}^{\mathrm{CFT}}\ket{\psi} = -\frac{c}{24\pi} + \frac{h}{\pi}-\frac{12h^{2}}{\pi}\frac{(2h-1)}{(4h-1)}\frac{1}{c} \,.
\end{equation}
The ADM mass or, equivalently, the energy of the state $\ket{\psi}_{\mathrm{CFT}}$ is
\begin{equation}\label{eq:ADMmass}
M = \int d\varphi\, \bra{\psi}T_{tt}^{\mathrm{CFT}}\ket{\psi} = -\frac{c}{12} + 2h - 24h^{2}\frac{(2h-1)}{(4h-1)}\frac{1}{c} \,.
\end{equation}
The leading term is the Casimir energy of the cylinder and the first subleading term is the bare conformal dimension of $O$. We have no choice but to interpret the $\mathcal{O}(c^{-1})$ term as giving rise to an anomalous dimension for the single-trace operator $O$:
\begin{equation}\label{eq:confdimcorrection}
\delta h = -12h^{2}\frac{(2h-1)}{(4h-1)} \,.
\end{equation}
Observe that the $\mathcal{O}(c^{-1})$ terms in the expectation value of the scalar field Hamiltonian and the ADM mass differ by a factor of $2$, which hints at a more general relation between the two. It would be interesting to understand this fact better, but we leave it for future work.

\subsubsection*{Anomalous dimension, mass renormalization and the bootstrap}

Using the bulk effective field theory and Einstein's equations, we have obtained the anomalous dimension of the single-trace operator in the CFT. To the best of our knowledge, this is the first computation of this type and offers a method similar to that of \cite{Fitzpatrick:2011hh} but for single-trace operators.

It is worthwhile to comment on the meaning of this anomalous dimension. In particular, it is important to emphasize again that we only have a bulk computation of $\delta h$, and that it is not fixed by first principles in the CFT. The large $N$ bootstrap is very much a bottom-up set up, in that we assume a certain set of starting conditions which can be viewed as input data (large $N$ factorization, large gap, etc) and derive constraints from the combination of these assumptions, along with crossing symmetry, unitarity and causality. In this setup, the scaling dimension of the single-trace operator is input data. Therefore, not only can we not verify the value of $\delta h$ as found from the bulk, but in fact it seems we are even allowed to force $\delta h=0$ in the CFT. How can that be compatible with the computation done in the bulk?

There is one more piece of the bulk EFT data that we have not discussed so far: the bulk mass. Nothing prevents us from expanding the mass itself in powers of $G_N$
\be
m=m^{(\textrm{bare})}+G_N m^{(1)} + \cdots
\ee
In this section, we have phrased everything in terms of $h$ using \rref{eq:mass-dimension}, so this would induce a shift
\be
h\to h + \delta h ^{\textrm{mass shift}} \,.
\ee
Using this, we could easily achieve a state whose energy is $2h$ up to order $c^{-2}$, simply by setting
\be
\delta h ^{\textrm{mass shift}}=-\delta h \,.
\ee
Therefore, the procedure should be seen as follows: first fix the scaling dimension of the CFT operator to the relevant order in the $1/c$ expansion (again, this is input data). Take into account the effect of backreaction, and tune $\delta h ^{\textrm{mass shift}}$ to match the CFT answer. This will always yield results that are compatible between the bulk and boundary, but the dictionary between bulk mass and CFT scaling dimension \rref{eq:mass-dimension} gets modified.

For the rest of the paper, since we have no a priori bias to work with protected operators, we will simply take the bulk expression $\delta h$ to be defining for us the CFT scaling dimension, i.e., we take $\delta h ^{\textrm{mass shift}}=0$. As a final comment, note that if the operator is BPS, the backreaction may not induce any correction to the energy, and the contribution of the bulk gauge field should cancel against the geometric backreaction, much like it does for double-trace operators in \cite{Fitzpatrick:2011hh}. It would be nice to check this explicitly.

%~~~~~~~~~~~~~~~~~~~~~~~~~~~~~~~~~~~~~~~~~~~~~~~
\section{Area operator}
\label{sec:Area} 
%~~~~~~~~~~~~~~~~~~~~~~~~~~~~~~~~~~~~~~~~~~~~~~

We now have all the pieces we need to compute the expectation value of the area operator to second-order (ignoring the contribution of the bulk entanglement entropy, as we will explain). Let us start by expanding the extremization of the generalized entropy appearing on the righthand side of \eqref{eq:HRRT} order-by-order in $G_{N}$. The metric and scalar field have already been expanded around pure AdS$_{3}$ in \eqref{eq:metricexp} and \eqref{eq:radialcompfreq_1storder}, respectively, and one can similarly expand the location of the quantum extremal surface around the classical extremal surface $\Sigma_{A}^{(0)}$:
\begin{equation}\label{eq:QESexp}
\Sigma_{A} = \Sigma_{A}^{(0)} + G_{N}\Sigma_{A}^{(1)} +G_{N}^{2} \Sigma_{A}^{(2)} + \ldots
\end{equation}
We now want to compute the difference of generalized entropies between the excited state and the vacuum. At zeroth order, one simply finds that the areas of the vacuum and the excited state cancel
\begin{equation}\label{eq:Sgenexp_minus1order}
\frac{1}{4G_N}\bra{0}\hat{A}[\Sigma_{A}^{(0)},g^{(0)}]\ket{0}-\frac{1}{4G_N}\bra{\psi^{(0)}}\hat{A}[\Sigma_{A}^{(0)},g^{(0)}]\ket{\psi^{(0)}}=0 \,.
\end{equation}
This is the piece that would be $\mathcal{O}(G_N^{-1})$ and is (CFT) UV-divergent in the vacuum. We see that it exactly cancels, in agreement with the CFT, which is both UV-finite and starts at order $c^0$. 

At first subleading order, the generalized entropy reads
\begin{equation}\label{eq:Sgenexp_0thorder}
\underset{\Sigma_A}{\ext}\left(\frac{\bra{\psi}\hat{A}[\Sigma_{A},g]\ket{\psi}}{4G_{N}}+S_{\mathrm{EE}}^{\mathrm{bulk}}[\Sigma_{A},g,\phi]\right)\Bigg|_{\mathcal{O}(G_{N}^{0})} = \frac{1}{4}\bra{\psi^{(0)}}\hat{A}[\Sigma_{A}^{(0)},g^{(1)}]\ket{\psi^{(0)}}+S_{\mathrm{EE}}^{\mathrm{bulk}}[\Sigma_{A}^{(0)},g^{(0)},\phi^{(0)}] \,,
\end{equation}
where no extremization is required at this order, that is, $\bra{\psi^{(0)}}\hat{A}[\Sigma_{A}^{(1)},g^{(0)}]\ket{\psi^{(0)}} = 0$ due to the classical extremality of $\Sigma_{A}^{(0)}$. This was computed in \cite{Belin:2018juv} and found to agree exactly with the CFT answer. At first order in $G_N$, one finds
\begin{equation}\label{eq:Sgenexp_1storder}
\begin{split}
\underset{\Sigma_{A}}{\ext}\Bigg(&\frac{\bra{\psi}\hat{A}[\Sigma_{A},g]\ket{\psi}}{4G_{N}}+S_{\mathrm{EE}}^{\mathrm{bulk}}[\Sigma_{A},g,\phi]\Bigg)\Bigg|_{\mathcal{O}(G_{N}^{1})} \\
	&= \Bigg[\frac{1}{4}\bra{\psi}\hat{A}[\Sigma_{A}^{(0)},g^{(2)}]\ket{\psi}+S_{\mathrm{EE}}^{\mathrm{bulk}}[\Sigma_{A}^{(0)},g^{(1)},\phi^{(0)}]+S_{\mathrm{EE}}^{\mathrm{bulk}}[\Sigma_{A}^{(0)},g^{(0)},\phi^{(1)}] \\
	&+\underset{\Sigma_{A}^{(1)}}{\ext}\bigg(\frac{1}{4}\bra{\psi^{(0)}}\hat{A}[\Sigma_{A}^{(1)},g^{(0)}]\ket{\psi^{(0)}}+\frac{1}{4}\bra{\psi^{(0)}}\hat{A}[\Sigma_{A}^{(1)},g^{(1)}]\ket{\psi^{(0)}}+S_{\mathrm{EE}}^{\mathrm{bulk}}[\Sigma_{A}^{(1)},g^{(0)},\phi^{(0)}]\bigg)\Bigg]G_{N} \,.
\end{split}
\end{equation}

Some comments are in order about this expansion:
\begin{itemize}
\item First, we have not written down the term $\bra{\psi^{(0)}}\hat{A}[\Sigma_{A}^{(2)},g^{(0)}]\ket{\psi^{(0)}} $ since it vanishes by extremality of $\Sigma_A^{(0)}$.
\item
The term $\bra{\psi}\hat{A}[\Sigma_{A}^{(0)},g^{(2)}]\ket{\psi}$ should be understood as capturing all $\mathcal{O}(G_{N}^{2})$ contributions to the area of the classical extremal surface $\Sigma_{A}^{(0)}$ coming from changes in the metric so it includes contributions to the area from $(g^{(1)})^{2}$ and from $g^{(2)}$.

\item
The term $S_{\mathrm{EE}}^{\mathrm{bulk}}[\Sigma_{A}^{(0)},g^{(0)},\phi^{(1)}]$ corresponds to all corrections to the bulk entanglement entropy from changes in the scalar field theory due to interactions with gravity, i.e., binding energies of two-particle states, corrections to coupling constants, etc.

\item The term $\bra{\psi^{(0)}}\hat{A}[\Sigma_{A}^{(1)},g^{(0)}]\ket{\psi^{(0)}}$ should be understood as capturing the quadratric dependence on $\Sigma_A^{(1)}$.

\end{itemize}

In this section, we will compute all the contributions to the area appearing in \eqref{eq:Sgenexp_1storder}, except for the contribution from the shift in the surface due to the bulk entanglement entropy appearing inside the extremization. Somewhat surprisingly, these geometric and entanglement contributions to the shape variation of the surface $\Sigma_{A}$ can be separated and computed independently, as we shall see in \S\ref{sec:shapevar}. This allows us to compute only the former in this paper.

%~~~~~~~~~~~~~~~~~~~~~~~~~~~~~~~~~~~~~~~~~~~~~~~
\subsection{Metric perturbation}
\label{sec:metricpert} 
%~~~~~~~~~~~~~~~~~~~~~~~~~~~~~~~~~~~~~~~~~~~~~~

We want to find the changes in the area of the quantum extremal surface due to perturbations of the metric. Define the induced metric on the surface $\Sigma_A$ by $h$, given the metric $g$ in the full spacetime. The expectation value of the area on surfaces $\Sigma_{A}$ homologous to $A$ is
\begin{equation}\label{eq:area}
\bra{\psi}\hat{A}[\Sigma_{A},g]\ket{\psi} = \int_{\Sigma_{A}}\sqrt{h} \,.
\end{equation}
The induced metric is
\begin{equation}\label{eq:inducedmetric}
h_{rr} = g_{rr}+\bigg(\frac{\partial \varphi}{\partial r}\bigg)^2g_{\varphi\varphi} \,.
\end{equation}
Extremizing the area functional with respect to the zeroth order induced metric $h^{(0)}$ gives the classical extremal surface described by
\begin{equation}\label{eq:classext}
\varphi'(r) = \frac{r_{\mathrm{min}}}{r\sqrt{(r^2+1)(r^2-r_{\mathrm{min}}^2)}} \,.
\end{equation}
where $r_{\mathrm{min}}$ is the deepest point reached in the bulk by the surface, which is related to $\theta$ by
\begin{equation}\label{eq:rmin}
r_{\mathrm{min}} = \cot\left(\frac{\theta}{2}\right) \,.
\end{equation}
The change in the area to second order in $G_{N}$ due to the change in the metric can now be computed from \eqref{eq:metric_1storder} and \eqref{eq:grr_2ndorder} and one finds
\begin{equation}\label{eq:areametricpert_2ndorder}
\bra{\psi}\hat{A}[\Sigma_{A}^{(0)},g^{(2)}]\ket{\psi} = G_{N}^2\int_{r_\mathrm{min}}^{\infty}dr\,\frac{\sqrt{1-\frac{r_{\mathrm{min}}^2}{r^2}}}{(r^2+1)^{\frac{3}{2}}}\left[\frac{64h^{2}}{r^{2}+1}\left(1-\frac{1}{(r^2+1)^{2h-1}}\right)^{2}\left(2+\frac{r_{\mathrm{min}}^{2}}{r^{2}}-r^{2}\right)-2G_{2}^{(2)}(r)\right] \,.
\end{equation}
This integral is difficult to evaluate exactly due to the complicated nature of $G_{2}^{(2)}$ (see \eqref{eq:G22def}), but it can be evaluated in the small interval (large $r_{\mathrm{min}}$) limit. The details are provided in App. \ref{sec:areacalcs} and here we simply state the result:
\begin{equation}
\label{eq:areametricpert_2ndorder_final}
\begin{split}
\frac{1}{4}\bra{\psi}&\hat{A}[\Sigma_{A}^{(0)},g^{(2)}]\ket{\psi} \\
	&= 8h^{2}G_{N}^{2}\Bigg[\frac{1}{(4h-1)}\theta^{2}\left({-}\frac{(2h-1)}{3}+\frac{2(143h-34)}{315}\left(\frac{\theta}{2}\right)^2+\mathcal{O}(\theta^4)\right) \\
	&+2\frac{\Gamma(\frac{3}{2})\Gamma(2h)}{\Gamma(2h+\frac{3}{2})}\left((4h^{2}-3h+1)-\frac{2h^{2}}{(4h-1)}\left(\psi(2h)-\psi(4h)\right)\right)\left(\frac{\theta}{2}\right)^{4h}\left(1+\mathcal{O}(\theta^2)\right) \\
	&+\frac{\Gamma(\frac{3}{2})\Gamma(4h-1)}{4\,\Gamma(4h+\frac{5}{2})}\left(\frac{\theta}{2}\right)^{8h-2}\left({-}(8h+1)(7h+3)+\left(96h^{3}+196h^{2}-h-13\right)\left(\frac{\theta}{2}\right)^2+\mathcal{O}(\theta^4)\right)\Bigg] \,.
\end{split}
\end{equation}
%

%~~~~~~~~~~~~~~~~~~~~~~~~~~~~~~~~~~~~~~~~~~~~~~~
\subsection{Shape variation}
\label{sec:shapevar} 
%~~~~~~~~~~~~~~~~~~~~~~~~~~~~~~~~~~~~~~~~~~~~~~

We next want to compute the change in the area due to the $G_{N}$ correction $\Sigma_{A}^{(1)}$ to the location of the extremal surface. This turns out to be simpler to compute in Rindler coordinates for reasons that will be explained momentarily.

We can define AdS-Rindler coordinates that cover the classical entanglement wedge $\mathcal{W}^{(0)}[A]$ defined by the bulk domain of dependence of $R_A^{(0)}$, the homology surface stretching between $A$ and $\Sigma_A^{(0)}$. The AdS-Rindler coordinates are $\tau,x \in \mathbb{R}$, $\rho \in [0,\infty)$ with coordinate transformation to global coordinates given by
\begin{equation}\label{eq:Rindlertoglobal}
\begin{split}
t &= \arctan\bigg(\frac{\sinh\rho\sinh\tau}{\sinh\rho\cosh\tau\sinh\eta+\cosh\rho\cosh x\cosh\eta}\bigg)
\\ r &= \sqrt{\cosh^{2}\rho\sinh^2 x + \big(\cosh\rho\cosh x\sinh\eta+\sinh\rho\cosh\tau\cosh\eta\big)^2}
\\ \varphi &= \arctan\bigg(\frac{\cosh\rho\sinh x}{\cosh\rho\cosh x\sinh\eta+\sinh\rho\cosh\tau\cosh\eta}\bigg) \,,
\end{split}
\end{equation}
where the boost parameter $\eta$ is related to the interval size by
\begin{equation}\label{eq:boost}
\eta = \cosh^{-1}\left(\csc\left(\frac{\theta}{2}\right)\right) \,.
\end{equation}
The AdS-Rindler metric is
\begin{equation}\label{eq:Rindlermetric}
(ds_{\mathrm{Rindler}}^{(0)})^{2} = -\sinh^{2}\rho \, d\tau^{2} + d\rho^2 + \cosh^{2}\rho \, dx^{2} \,.
\end{equation}
One particularly nice feature of these coordinates is that the classical extremal surface $\Sigma_{A}^{(0)}$ lies at $\rho=0$. Therefore, the quantum extremal surface at first order in $G_{N}$ away from $\Sigma_{A}^{(0)}$ is described by
\begin{equation}\label{eq:shapevar}
\rho(x) = G_{N}\rho^{(1)}(x) \,.
\end{equation}

Let us now examine the quantum extremal surface equation in these coordinates. The Lagrangian for the area is
\begin{equation}\label{eq:RindlerareaL}
\mathcal{L}_{A} = \sqrt{h} = \sqrt{g_{xx}+2\rho'(x)g_{x\rho}+g_{\rho\rho}(\rho'(x))^{2}} \,.
\end{equation}
We can expand $\mathcal{L}_{A}$ to second order in $G_{N}$, ignoring any $(g^{(1)})^2$ and $g^{(2)}$ terms since these contributions were computed in \S\ref{sec:metricpert}, and we find
\begin{equation}\label{eqn:RindlerareaL_2ndorder}
\mathcal{L}_{A}^{(2)} \supset \frac{1}{2}\left(({\rho^{(1)}}'(x))^{2}+(\rho^{(1)}(x))^2-V_{1}(x)\rho^{(1)}(x)-V_{2}(x){\rho^{(1)}}'(x)\right) \,,
\end{equation}
where we have defined the `potentials'
\begin{equation}\label{eq:V12}
\begin{split}
V_{1}(x) &= \frac{32h\sech x \tanh^{2}x \tanh \eta}{(\cosh x \cosh \eta)^{4h}(\sinh^{2}x+\cosh^{2}x \sinh^{2}\eta)^{2}} 
\\	& \times \Bigg[\cosh^{2}x\cosh^{2}\eta\left((\cosh x \cosh \eta)^{4h}-\cosh^{2}x\cosh^{2}\eta\right)
\\	&+\left((\cosh x \cosh \eta)^{4h}-2h\cosh^{2}x\cosh^{2}\eta\right)\left(\sinh^{2}x+\cosh^{2}x \sinh^{2}\eta\right)\Bigg]
\\ V_{2}(x) &= 32h\sech x \tanh x \tanh \eta \frac{(\cosh^{2}x \cosh^{2}\eta - (\cosh x \cosh \eta)^{4h})}{(\cosh x \cosh \eta)^{4h}(\sinh^{2}x+\cosh^{2}x \sinh^{2}\eta)} \,.
\end{split}
\end{equation}
Notice that we did not include any contribution from the second order variation of the surface $\rho^{(2)}(x)$ because classical extremality of $\Sigma_{A}^{(0)}$ implies that any such contribution is higher order in $G_{N}$.

The change in the bulk entanglement entropy due to the shape variation of the surface was derived via the path integral in \cite{Rosenhaus:2014woa}. They found that it is given by the following integral over Euclidean AdS$_{3}$:
\begin{equation}\label{eq:bulkEE_shapevar}
S_{\mathrm{EE}}^{\mathrm{bulk}}[\Sigma^{(1)},g^{(0)}] = \lim_{\epsilon \to 0}\frac{1}{2}\int_{\mathrm{EAdS}_{3}\backslash \mathcal{R}_{\epsilon}} d^{3}x\,\sqrt{g_{E}^{(0)}}\delta g_{E,\mu\nu}\bra{\psi^{(0)}} (T_{E})^{\mu\nu}K_{\psi} \ket{\psi^{(0)}}_{c} \,,
\end{equation}
where $\delta g_{E,\mu\nu}$ is a diffeomorphism that maps $\Sigma_{A}^{(0)}$ to $\Sigma_{A}$ and $K_{\psi} = -\log \rho_{\psi}$ is the modular Hamiltonian associated to the reduced density matrix for the state $\ket{\psi^{(0)}}$ in $\mathcal{W}^{(0)}[A]$. The subscript $c$ on the two-point function denotes the connected correlator. Futhermore, we have cut out a tubular neighborhood $\mathcal{R}_{\epsilon}$ of $\Sigma_{A}^{(0)}$ with radius $\epsilon$ because the integrand diverges there. After integrating by parts and using conservation of the stress-tensor, this reduces to
\begin{equation}\label{eq:bulkEE_shapedeform}
S_{\mathrm{EE}}^{\mathrm{bulk}}[\Sigma^{(1)},g^{(0)}] = G_{N}\lim_{\epsilon \to 0}2\pi\epsilon\int_{\Sigma_{A}^{(0)}} dx\,\sqrt{h_{E}^{(0)}}\rho^{(1)}(x)\bra{\psi^{(0)}} (T_{E})^{\rho\rho}(x,\rho=\epsilon)K_{\psi}\ket{\psi^{(0)}}_{c} \,.
\end{equation}
The Lagrangian for the bulk entanglement entropy at order $G_{N}$ thus takes the form
\begin{equation}
\mathcal{L}_{\mathrm{EE}} \equiv G_{N}\frac{V_{\mathrm{EE}}(x)}{4}\rho^{(1)}(x) \,.
\end{equation}
The total Lagrangian is
\begin{equation}\label{eq:Ltot}
\mathcal{L} = \frac{1}{4G_{N}}\mathcal{L}_{A} + \mathcal{L}_{\textrm{EE}}
\end{equation}
whose Euler-Lagrange equation at first order in $G_{N}$ is
\begin{equation}\label{eq:E-L}
{\rho^{(1)}}''(x) = \rho^{(1)}(x) + V_{\mathrm{geo}}(x) + V_{\textrm{EE}}(x) \,,
\end{equation}
where
\begin{equation}\label{eq:Vcldef}
V_{\mathrm{geo}}(x) \equiv \frac{1}{2}\left(V_{2}'(x)-V_{1}(x)\right) \,.
\end{equation}
This is a simple inhomogeneous second-order differential equation with solution
\begin{equation}\label{eq:E-Lsoln}
\rho^{(1)}(x) = \mathcal{A}_{+}e^{x}+\mathcal{A}_{-}e^{-x}+\frac{1}{2}\left(e^{x}\int dx\,e^{-x}\left(V_{\mathrm{geo}}(x) + V_{\textrm{EE}}(x)\right) - e^{-x}\int dx\,e^{x}\left(V_{\mathrm{geo}}(x) + V_{\textrm{EE}}(x)\right)\right) \,.
\end{equation}
The constants $\mathcal{A}_{\pm}$ are fixed by the boundary conditions $\rho^{(1)}(\pm\infty) = 0$. This demonstrates explicitly why we can separate the shift in the extremal surface into a geometry piece and a bulk entanglement piece. One should view the change in the geometry and the change in the bulk entanglement entropy as two forces that pull on the quantum extremal surface, each in their independent way. In particular, we can write
\begin{equation}\label{eq:shapedeform_decomp}
\rho^{(1)}(x) = \rho_{\mathrm{geo}}^{(1)}(x) + \rho_{\textrm{EE}}^{(1)}(x)
\end{equation}
with
\begin{equation}\label{eq:shapedeform_qcl}
\rho_{\mathrm{geo}/\textrm{EE}}^{(1)}(x) = \mathcal{A}_{+}^{\mathrm{geo}/\textrm{EE}}e^{x}+\mathcal{A}_{-}^{\mathrm{geo}/\textrm{EE}}e^{-x}+\frac{1}{2}\left(e^{x}\int dx\,e^{-x}V_{\mathrm{geo}/\textrm{EE}}(x)  - e^{-x}\int dx\,e^{x}V_{\mathrm{geo}/\textrm{EE}}(x)\right) \,.
\end{equation}
We can now ignore $\rho_{\textrm{EE}}^{(1)}(x)$ for the purposes of this paper and leave the calculation of this quantity to \cite{ourpapertoappear}. To compute $\rho_{\mathrm{geo}}^{(1)}(x)$, we perform the small interval expansion and find
\begin{equation}\label{eq:shapedeform_soln}
\begin{split}
\rho_{\mathrm{geo}}^{(1)}(x) &= \theta^{2}\frac{h}{3}(\cosh(2x)+3)\sech^{3}x\left(1+\mathcal{O}(\theta^{2})\right) + \left(\frac{\theta}{2}\right)^{4h}\left(1+\mathcal{O}(\theta^{2})\right)\Bigg[{-}8h\frac{\Gamma(2h+1)\Gamma(\frac{3}{2})}{\Gamma(2h+\frac{3}{2})}e^{x} \\
	&-4h\sinh x\sech^{4h+2}x\left(\frac{1}{(2h+1)}+\frac{e^{-2x}}{(h+1)}\left(1-{}_{2}{F}_{1}\left(1,-2h-2,2h+1;-e^{2x}\right)\right)\right)\Bigg] \,.	
\end{split}
\end{equation}

Performing the integral in $x$ of \eqref{eqn:RindlerareaL_2ndorder}, we obtain the change in the area of the quantum extremal surface due to the shape variation
\begin{equation}\label{eq:areashapevar_2ndorder}
\begin{split}
\frac{1}{4}\bra{\psi}\hat{A}[\Sigma_{A,\mathrm{geo}}^{(1)},g]\ket{\psi}|_{\mathcal{O}(G_{N}^{2})} &= -\frac{164}{315}h^{2}\theta^{4} + \frac{128h^{2}\,\Gamma(2h+1)\Gamma(\frac{3}{2})}{3\Gamma(2h+\frac{7}{2})}(8h^{2}+19h+9)\left(\frac{\theta}{2}\right)^{4h+2}\left(1+\mathcal{O}(\theta^{2})\right) \\
	&-32h^{2}\left(2\frac{\Gamma(\frac{3}{2})^{2}\Gamma(2h+1)^{2}}{\Gamma(2h+\frac{3}{2})^{2}}+\frac{\Gamma(\frac{3}{2})\Gamma(4h+1)}{\Gamma(4h+\frac{5}{2})}\right)\left(\frac{\theta}{2}\right)^{8h}\left(1+\mathcal{O}(\theta^{2})\right) \,.
\end{split}
\end{equation}
The details of the calculation of this integral can be found in App. \ref{sec:areacalcs}. 

Finally, all our results for the area to second order can be combined to obtain
\begin{equation}\label{eq:totalclassicalarea}
\begin{split}
\frac{1}{4G_{N}}\Big(\bra{\psi}\hat{A}&[\Sigma_{A,\mathrm{geo}}^{(1)},g]\ket{\psi}+\bra{\psi}\hat{A}[\Sigma_{A}^{(0)},g]\ket{\psi}\Big)\Big|_{\mathcal{O}(G_{N}^{2})} \\
	&= G_{N}\Bigg({-}\frac{2h^{2}}{3}\theta^{2}\left[\frac{4(2h-1)}{(4h-1)} + \left(1+\frac{(2h-1)}{(4h-1)}\right)\frac{1}{15}\theta^{2} + \mathcal{O}(\theta^{4})\right] \\
	&+16h^{2}\frac{\Gamma(\frac{3}{2})\Gamma(2h)}{\Gamma(2h+\frac{3}{2})}\left((4h^{2}-3h+1)-\frac{2h^{2}}{(4h-1)}\left(\psi(2h)-\psi(4h)\right)\right)\left(\frac{\theta}{2}\right)^{4h}\left(1+\mathcal{O}(\theta^2)\right) \\
	&+2h^{2}\frac{\Gamma(\frac{3}{2})\Gamma(4h-1)}{\Gamma(4h+\frac{5}{2})}\left(\frac{\theta}{2}\right)^{8h-2}\Bigg({-}(8h+1)(7h+3) \\
	&+\left(96h^{3}-60h^{2}+63h-13-32\frac{\Gamma(\frac{3}{2})\Gamma(2h+1)^{2}\Gamma(4h+\frac{5}{2})}{\Gamma(2h+\frac{3}{2})^{2}\Gamma(4h-1)}\right)\left(\frac{\theta}{2}\right)^2+\mathcal{O}(\theta^4)\Bigg) \,.
\end{split}
\end{equation}
Focusing only on the $\theta^2$ and $\theta^4$ terms, we can rewrite the answer using $c=\frac{3}{2G_N}$ and $\delta h =-12h^{2}\frac{(2h-1)}{(4h-1)}$ to obtain
\begin{equation}\label{eq:totalclassicalarea_thetan}
\begin{split}
\frac{1}{4G_{N}}\Big(\bra{\psi^{(0)}}\hat{A}[\Sigma_{A,\mathrm{geo}}^{(1)},g]\ket{\psi^{(0)}}&+\bra{\psi}\hat{A}[\Sigma_{A}^{(0)},g]\ket{\psi}\Big)\big|_{\mathcal{O}(G_{N}^{2}),\mathcal{O}(\theta^2)+\mathcal{O}(\theta^4)} \\
	&= \frac{2\delta h}{c}\left(\frac{\theta^2}{6}+\frac{\theta^4}{360}\right) -\frac{16h^2}{15c}\left( \frac{\theta}{2} \right)^{4} \,.
\end{split}
\end{equation}
We see that the $\theta^{2}$ and $\theta^{4}$ terms above exactly agree with the CFT answer found in \eqref{eq:CFTresult_2ndorder} once expanded, thus confirming the first row of the $\mathcal{O}(c^{-1})$ part of the dictionary in Table \ref{tab:dict}.

%~~~~~~~~~~~~~~~~~~~~~~~~~~~~~~~~~~~~~~~~~~~~~~~
\section{Discussion}
\label{sec:discussion}
%~~~~~~~~~~~~~~~~~~~~~~~~~~~~~~~~~~~~~~~~~~~~~~

In this paper, we have computed the entanglement entropy in a holographic CFT$_2$ for one interval in a state obtained by acting with a primary single-trace operator on the vacuum. We performed the computation in a double expansion in $1/c$ and in the interval size, and obtained results to order $c^{-1}$. In the bulk, the state maps to a perturbative one-particle state of the bulk matter, working at an order that goes beyond the FLM formula and where quantum extremality becomes important. We computed the expectation value of the extremal area operator, considering all effects but the displacement of the surface induced by the bulk entanglement. We found that all CFT terms involving the Virasoro identity block were fully accounted for by this area operator, while the contributions of double-trace operators must be encoded in the bulk entanglement entropy. We now conclude with open questions and future directions.

\subsection{The bulk entanglement entropy, and what is left to be done}

In this paper, we have computed the expectation value of the area operator, taking into account all effects but the variation of the surface due to the ``pull'' by the entanglement entropy. We left the evaluation of the bulk entanglement entropy to \cite{ourpapertoappear}. We now comment on the remaining pieces that need to be evaluated. There are three contributions that need to be computed as detailed below
\begin{enumerate}
\item First, we need to evaluate $S_{\textrm{EE}}^{\mathrm{bulk}}[\Sigma_{A}^{(0)},g^{(1)},\phi^{(0)}]$. This term is not conceptually difficult, it computes the change in the entanglement entropy due to the change in the background geometry. It is given by a bulk two-point function between the stress-tensor and the modular Hamiltonian \cite{Rosenhaus:2014woa}. Since the bulk theory is free, and the modular Hamiltonian of the Rindler-wedge is local, this can be computed explicitly.
\item The second term consists of the shape variation involving $\rho_{\textrm{EE}}^{(1)}$, which appears in the extremization of the quantum extremal surface. As with metric variations, the change in the entanglement entropy is computed by a two-point function between the stress-tensor and the modular Hamiltonian \cite{Rosenhaus:2014woa}, as reviewed in \eqref{eq:bulkEE_shapevar} and \eqref{eq:bulkEE_shapedeform}. Like the previous term, this can be computed explicitly and is not conceptually difficult. However, both in this term and in the previous one, there could be bulk UV-divergences to worry about, which would be a new feature compared to \cite{Belin:2018juv} where everything was bulk UV-finite.
\item Finally, there is the term $S_{\textrm{EE}}^{\mathrm{bulk}}[\Sigma_{A}^{(0)},g^{(0)},\phi^{(1)}]$ which encodes the fact that the scalar field sector is no longer free due to its interactions with gravity. This term will involve breaking the Fock structure of the Hilbert space, and can be tackled using the technology of \cite{Fitzpatrick:2011hh} that ``integrates out" gravitons. Note also that this term would be the most relevant term if we had considered a $\lambda \phi^4$ interaction in the bulk.
\end{enumerate}

\subsection{Bulk cancellations}

We have seen that the (classical) area operator gives us entirely the exchanges of the identity operator and the stress-tensor in the CFT. However, it also gives us many other terms, some that scale like $\theta^{4h}$ and some that scale like $\theta^{8h}$. We know for a fact that the $\theta^{4h}$ terms must cancel against other contributions in the bulk entanglement entropy, since such a term is absent from the CFT answer. This could be understood better by computing the relative entropies \`a  la JLMS \cite{Jafferis:2015del} to this order, and would involve modular extremal surfaces \cite{Dong:2017xht}. These cancellations would presumably be a more involved version of the bulk first law of entanglement found in \cite{Belin:2018juv}. 

The $\theta^{8h}$ term on the other hand is harder to decode. At this stage, it could give a contribution to the final answer that does not cancel against anything else in the bulk entanglement entropy, since a $\theta^{8h}$ term also appears in the CFT answer. It would then correspond to part of the modification to the OPE coefficients of double-trace operators \rref{eq:CFTanswer_DTOPE}. The coefficients do not appear to match, so it cannot make up for this alone, and would in any case need to be accompanied by some other contribution from the bulk entanglement entropy. It is also possible that it completely cancels against the bulk entanglement entropy, meaning that the area operator accounts for the Virasoro block alone, up to things that cancel against the bulk entanglement entropy. It would be interesting to understand this better, and it will be explored in \cite{ourpapertoappear}.

Taking a step back, it is interesting to think about these terms from the CFT point of view. Since they do not appear in any CFT quantity, it is interesting to ask what their meaning is. These terms come from separating the bulk entanglement entropy and the area operator. Since only the combination is gauge-invariant and bulk UV-finite, is it possible that these terms are not true gauge-invariant quantities? Everything we have computed is clearly gauge-invariantly defined in the bulk, so they would have to be not gauge-invariant in some generalized sense. Or if we take the perspective that these terms are meaningful on their own, how do we extract them from the CFT? We hope to return to these questions in the future.

\subsection{Graviton entanglement}

It is also worth discussing the entanglement of (boundary) gravitons. In everything we have discussed, we have not considered the possibility that the scalar state polarizes the entanglement structure of boundary gravitons. Since the scalar and graviton sectors interact at order $1/c$, one may think that inserting the scalar operator is not very different from inserting a stress-tensor, and the insertion of a stress-tensor would certainly polarize the graviton entanglement (just as inserting a $U(1)$ current polarizes the photon entanglement \cite{Belin:2019mlt}).

While we cannot prove this without also computing the bulk entanglement entropy for the scalar and matching to the CFT answer, we do not believe this to be the case. Since the graviton sector is topological in AdS$_3$, inserting a non-trivial scalar state does not affect this topological sector, and the local modification of the geometry obviously doesn't change the topology of the entangling surface. So we believe that at this order, it is not necessary to consider the graviton entanglement since the contribution will be the same as that of the vacuum and cancel in the difference of entanglement entropies. It is of course possible that our intuition is wrong, or that this statement fails at higher order in the $1/c$ expansion. But with the information we currently have, it is natural to conjecture that
\begin{equation}\label{eq:Virasoro}
\text{Virasoro Id block} \subset \frac{A_{\ext}}{4G_N} \,, \qquad \text{to all orders in $1/c$} \,,
\end{equation}
namely that the Virasoro identity block is fully captured by the area operator (up to terms that cancel against the bulk entanglement entropy) and that the entanglement of gravitons does not play any role to any order in the $1/c$ expansion.

There are two cases where we do expect the boundary gravitons and their entanglement structure to play an important role: the first is to consider a state where we insert a Virasoro descendant of the primary operator. This will certainly polarize the entanglement structure of boundary gravitons, much like inserting a stress-tensor does. The second is to go to higher dimensions, where the gravitational sector is no longer topological. There, we do expect the entanglement of gravitons to play an important role. We hope to return to this in the future.

\subsection{Excited states vs multiple intervals}

Finally, we comment on a bootstrap-type approach for multiple intervals and its difference with excited states. Trying to understand the conditions under which one obtains a HRRT formula for multiple intervals in the vacuum was a task undertaken in \cite{Headrick:2010zt,Hartman:2013mia,Belin:2017nze}. This type of approach more naturally connects to the modular bootstrap program, since it involves constraining torus or higher genus partition functions. Demanding that the leading $\mathcal{O}(c)$ term matches with the bulk HRRT formula leads to conditions on the spectrum \`a la HKS \cite{Hartman:2014oaa}, or generalizations for OPE coefficients \cite{Belin:2017nze}. A first downside is that these conditions typically do not force the CFT to be holographic, and a small $\Delta_{\textrm{gap}}$ is allowed.

Computing the quantum corrections is typically hard, which in the bulk are given by 1-loop determinants on the handle-body geometries relevant for the \ren entropies \cite{Faulkner:2013yia,Barrella:2013wja}. We are not aware of any computation that probes $\mathcal{O}(c^{-1})$ effects where quantum extremality becomes important, but our expectations are that again these corrections are given by heat-kernels of the (now interacting) bulk perturbative fields. The important point is that we do not expect the \textit{background geometry} to change at all, and we believe the handlebody geometries (which are locally AdS) will still accurately describe the state. In Lorentzian signature, these geometries have horizons, surfaces that become the HRRT surfaces as $n\to1$ and we do not expect these surfaces to move, in principle to any order in the $1/c$ expansion. Therefore, it seems that the structure of the quantum corrections will be less rich than in our setup where we explicitly saw the surface move. It would be interesting to understand this better.

%~~~~~~~~~~~~~~~~~~~~~~~~~~~~~~~~~~~~~~~~~~~~~~

%~~~~~~~~~~~~~~~~~~~~~~~~~~~~~~~~~~~~~~~~~~~~~~~
\acknowledgments
%~~~~~~~~~~~~~~~~~~~~~~~~~~~~~~~~~~~~~~~~~~~~~~

It is a pleasure to thank A. Castro, J. de Boer, N. Engelhardt, L. Fitzpatrick, T. Hartman, N. Iqbal, M. Meineri, R. Myers, K. Papadodimas, G. Sarosi, D. Turton, M. Walters for useful discussions. We would like to thank KITP, UCSB for hospitality during the workshop ``Gravitational Holography'' where this work was initiated. AB is grateful to the participants of the
SwissMap workshop ``Advances in Quantum Gravity" for stimulating discussions. This research was supported in part by the National Science Foundation under Grant No. NSF PHY-1748958. SCE was supported by U.S. Department of Energy grant DESC0019480 under the HEP-QIS QuantISED program.

\appendix

%~~~~~~~~~~~~~~~~~~~~~~~~~~~~~~~~~~~~~~~~~~~~~~~
\section{Bulk wavefunction and metric backreaction}
\label{sec:bulkwfn+metricbackreact}
%~~~~~~~~~~~~~~~~~~~~~~~~~~~~~~~~~~~~~~~~~~~~~~~

In this appendix, we provide the details of the calculation of the second-order backreaction of the scalar field on the metric discussed in \S\ref{sec:BulkEFT}. 

%~~~~~~~~~~~~~~~~~~~~~~~~~~~~~~~~~~~~~~~~~~~~~~~
\subsection{Wavefunction}
\label{sec:bulkwfn}
%~~~~~~~~~~~~~~~~~~~~~~~~~~~~~~~~~~~~~~~~~~~~~~~

We start by solving the scalar wave equation \eqref{eq:waveeqn_1storder} on the first-order backreacted spacetime \eqref{eq:metric_1storder}. It is easiest to do this in $\alpha$ coordinates with $r = \tan\alpha$. Furthermore, we can simplify our formulae by working with a redefined perturbative parameter
\begin{equation}\label{eq:epsilondef}
\epsilon \equiv 16hG_{N}.
\end{equation}
The Laplace-Beltrami operator on the first-order backreacted spacetime is
\begin{equation}\label{eq:Laplaceop}
\nabla^2 = \frac{1}{\sqrt{-g}}\partial_{\mu}\left(\sqrt{-g}\,g^{\mu\nu}\partial_{\nu}\right) = \big(\nabla^{(0)}\big)^2+\epsilon D^{(1)}+\mathcal{O}(\epsilon^2) \,,
\end{equation}
where
\begin{equation}\label{eq:Laplaceop_defs}
\begin{split}
\big(\nabla^{(0)}\big)^2 &= -\cos^2\alpha\,\partial_{t}^2+\cot\alpha\,\partial_{\alpha} + \cos^2\alpha\,\partial_{\alpha}^2 + \cot^{2}\alpha\,\partial_{\varphi}^{2}+\mathcal{O}(\epsilon^2) \\
D^{(1)} &\equiv -\cos^{4}\alpha\,\partial_{t}^2+\left({-}\cos^{2}\alpha\cot\alpha\cos(2\alpha)+\cos^{4h+1}\alpha\csc\alpha\left(h\cos(2\alpha)-h+1\right)\right)\partial_{\alpha} \\
	&+\cos^{4}\alpha\left(\cos^{4h-2}\alpha-1\right)\partial_{\alpha}^{2} \,.
\end{split}
\end{equation}
Expanding the wavefunction $f_{0,0}(\alpha)$ around the pure AdS$_{3}$ wavefunction as in \eqref{eq:radialcompfreq_1storder}, the wave equation at first order in $\epsilon$ becomes
\begin{equation}\label{eq:waveeq_explicit}
\left(\left(\nabla^{(0)}\right)^2-m^2\right)e^{-2iht}f_{0,0}^{(1)}(\alpha) = -\left(D^{(1)}+4h\cos^2\alpha\,\Omega_{0,0}^{(1)}\right)e^{-2iht}f_{0,0}^{(0)}(\alpha) \,.
\end{equation}
We make one more change of coordinates
\begin{equation}\label{eq:ycoords}
f_{0,0}^{(1)}(\alpha) = \cos^{2h}\alpha\, g^{(1)}(y), \qquad y = \sin^2\alpha \,,
\end{equation}
so that \eqref{eq:waveeq_explicit} becomes
\begin{equation}\label{eq:g1ode_y}
\left(\partial_{y}^2+\frac{(1-(1+2h)y)}{y(1-y)}\partial_{y}\right)g^{(1)}(y) = Q(y) \,,
\end{equation}
where
\begin{equation}\label{eq:Qdef}
Q(y) \equiv -\frac{1}{\sqrt{2\pi}}\frac{1}{y(1-y)}\left(h^{2}(1-y)+h\Omega_{0,0}^{(1)}-h(1-y)^{2h-1}(1-2hy)+h\left(1-(1+h)y\right)\right) \,.
\end{equation}
This is a second-order linear inhomogeneous differential equation. To solve this, let us first look at the homogeneous differential equation:
\begin{equation}\label{eq:homog}
\left(\partial_{y}^2+\frac{(1-(1+2h)y)}{y(1-y)}\partial_{y}\right)P(y) = 0 \,.
\end{equation}
This is just the pure AdS wave equation for $n=m=0$ with the two solutions:
\begin{equation}\label{eq:homogsolns}
P_{1}(y) = 1, \qquad P_{2}(y) = \ln y +2hy \,{}_{3}{F}_{2}\left(1,1,2h+1;2,2;y\right) \,.
\end{equation}
The Wronskian is given by
\begin{equation}\label{eq:Wronsk}
W(y) = P_{1}(y)P_{2}'(y)-P_{2}(y)P_{1}'(y) = \frac{1}{y(1-y)^{2h}} \,.
\end{equation}
The inhomogeneous differential equation \eqref{eq:g1ode_y} thus has the following solution:
\begin{equation}\label{eq:g1soln_gen}
g^{(1)}(y) = \mathcal{C}_{1}P_{1}(y)+\mathcal{C}_{2}P_{2}(y)+g_{P}(y) \,,
\end{equation}
where
\begin{equation}\label{eq:gP}
g_{P}(y) = P_{2}(y)\int dy\,\frac{P_{1}(y)}{W(y)}Q(y)-P_{1}(y)\int dy\,\frac{P_{2}(y)}{W(y)}Q(y) \,,.
\end{equation}
After integration by parts, we find
\begin{equation}\label{eq:gP_final}
\begin{split}
g_{P}(y) &= -\frac{h}{2\sqrt{2\pi}}\int dy\,\frac{1}{y(1-y)^{2h}}\left[\frac{1}{(4h-1)}(1-y)^{4h-1}\left(1-(4h-1)y\right)+(1-y)^{2h}\left(2y-1-\frac{\Omega_{0,0}^{(1)}}{h}\right)\right] \\
	&= -\frac{h}{2\sqrt{2\pi}}\left[\frac{1}{2h}(1-y)^{2h}-\frac{(2h-1)}{(4h-1)}y \,{}_{3}{F}_{2}\left(1,1,2-2h;2,2;y\right)+2y-\left(1+\frac{\Omega_{0,0}^{(1)}}{h}-\frac{1}{(4h-1)}\right)\ln y\right] \,.
\end{split}
\end{equation}
Putting all of the pieces together, the first-order radial wavefunction is
\begin{equation}\label{eq:f1}
\begin{split}
f_{0,0}^{(1)}(y) &= (1-y)^{h}\Bigg(\mathcal{C}_{1}+\mathcal{C}_{2}\left(\ln y + 2hy \,{}_{3}{F}_{2}\left(1,1,2h+1;2,2;y\right)\right) \\
	&-\frac{h}{2\sqrt{2\pi}}\left[\frac{1}{2h}(1-y)^{2h}-\frac{(2h-1)}{(4h-1)}y \,{}_{3}{F}_{2}\left(1,1,2-2h;2,2;y\right)+2y-\left(1+\frac{\Omega_{0,0}^{(1)}}{h}-\frac{1}{(4h-1)}\right)\ln y\right]\Bigg) \,.
\end{split}
\end{equation}

It remains to determine the constants $\mathcal{C}_{1}$, $\mathcal{C}_{2}$, and $\Omega_{0,0}^{(1)}$. First, we require that the wavefunction be regular at the origin ($y=0$) which fixes
\begin{equation}\label{eq:C2}
\mathcal{C}_{2} = -\frac{h}{2\sqrt{2\pi}}\left(1+\frac{\Omega_{0,0}^{(1)}}{h}-\frac{1}{(4h-1)}\right) \,.
\end{equation}
Next, we require normalizability of the wavefunction with respect to the Klein-Gordon inner product so the wavefunction must not diverge at the boundary ($y=1$). All terms in $f_{0,0}^{(1)}$ are finite at the boundary except ${}_{3}{F}_{2}\big(1,1,2h+1;2,2;y\big)$, and hence we must have
\begin{equation}\label{eq:Omega1}
\mathcal{C}_{2} = 0 \implies \Omega_{0,0}^{(1)} = -2h\frac{(2h-1)}{(4h-1)} \,.
\end{equation}
Finally, we choose to normalize our wavefunction to have unit norm with respect to the Klein-Gordon inner product defined by
\begin{equation}\label{eq:KG}
\langle \phi_{1},\phi_{2} \rangle = i\int_{\sigma} d^{2}x\sqrt{g}g^{tt}\left(\phi_{1}^{\ast}\partial_{t}\phi_{2}-\phi_{2}^{\ast}\partial_{t}\phi_{1}\right) \,,
\end{equation}
where $\sigma$ is a spacelike slice. The wavefunction $f_{0,0}^{(0)}$ is normalized with unit norm so $f_{0,0}^{(1)}$ must have vanishing norm at $\mathcal{O}(\epsilon)$, leading to
\begin{equation}\label{eq:zeronorm}
\begin{split}
0 &= \left[2\left(\Omega_{0,0}^{(0)}+\epsilon\Omega_{0,0}^{(1)}\right)\int dy\,d\varphi\,\sqrt{-\widetilde{g}}\,\widetilde{g}^{tt}\widetilde{f}_{0,0}(y,\varphi)^{2}\right]\bigg|_{\mathcal{O}(\epsilon)} \\
	&= \frac{1}{2}\int_{0}^{1} \frac{dy}{(1-y)}\left[\left(\Omega_{0,0}^{(1)}+\Omega_{0,0}^{(0)}(1-y)-\frac{\Omega_{0,0}^{(0)}}{2}(1-y)^{h}\right)f_{0,0}^{(0)}(y)+2\Omega_{0,0}^{(0)}f_{0,0}^{(1)}(y)\right]f_{0,0}^{(0)}(y) \\
	&= -\frac{8h-3}{8\pi(4h-1)}+\frac{\mathcal{C}_{1}}{\sqrt{2\pi}}+\frac{h}{4\pi(4h-1)}\left(\psi(4h)-\psi(2h+1)\right) \\
	\implies \mathcal{C}_{1} &=  \frac{1}{2\sqrt{2\pi}}\left(1+\frac{h}{(4h-1)}\left(\psi(2h)-\psi(4h)\right)\right) \,,
\end{split}
\end{equation}
where the integral of the hypergeometric ${}_{3}{F}_{2}$ can be performed using 7.512.11 in \cite{GradRyz}. Therefore, we obtain the first-order radial wavefunction appearing in \eqref{eq:wavefn_1storder_final}.

%~~~~~~~~~~~~~~~~~~~~~~~~~~~~~~~~~~~~~~~~~~~~~~~
\subsection{Second-order metric}
\label{sec:metric_2ndorder}
%~~~~~~~~~~~~~~~~~~~~~~~~~~~~~~~~~~~~~~~~~~~~~~~

We next explain how to solve Einstein's equations to obtain the second-order backreacted metric. We make the following metric ansatz for our asymptotically AdS$_{3}$ spacetime in global coordinates:
\begin{equation}\label{eqn:AAdS3metric}
ds^2 = -F_{1}(r)dt^2+\frac{dr^2}{F_{2}(r)}+r^2d\varphi^2 \,,
\end{equation}
where
\begin{equation}\label{eq:F12}
F_{1,2}(r) = r^2+G_{1,2}(r)^2 \,.
\end{equation}
We expand the functions $G_{1,2}(r)$ around pure AdS as
\begin{equation}\label{eq:G12}
G_{1,2}(r) = 1+G_{N}G_{1,2}^{(1)}(r)+G_{N}^2G_{1,2}^{(2)}(r) \,,
\end{equation}
The Ricci tensor and Ricci scalar are found to be
\begin{equation}\label{eq:Ricci}
\begin{split}
R_{tt} &= \frac{1}{4}F_{1}'F_{2}'+\frac{1}{2}F_{2}F_{1}''-\frac{1}{4}\frac{F_{2}(F_{1}')^2}{F_{1}}+\frac{1}{2}\frac{F_{2}F_{1}'}{r} \\
R_{rr} &= \frac{1}{4}\frac{(F_{1}')^2}{F_{1}^2}-\frac{1}{2}\frac{F_{1}''}{F_{1}}-\frac{1}{2}\frac{F_{2}'}{F_{2}r}-\frac{1}{4}\frac{F_{2}'F_{1}'}{F_{2}F_{1}} \\
R_{\varphi\varphi} &= -\frac{1}{2}rF_{2}'-\frac{1}{2}\frac{rF_{1}'F_{2}}{F_{1}} \\
R &= \frac{1}{2}\frac{(F_{1}')^2F_{2}}{F_{1}^2}-\frac{1}{2}\frac{F_{2}'F_{1}'}{F_{1}}-\frac{F_{2}F_{1}'}{rF_{1}}-\frac{F_{2}F_{1}''}{F_{1}}-\frac{F_{2}'}{r} \,.
\end{split}
\end{equation}
We can now write Einstein's equations in terms of $G_{1,2}(r)$. The $tt$ Einstein equation at second order \eqref{EEtt_2ndorder} becomes
\begin{equation}\label{eq:ttEE_explicit}
\begin{split}
-\left(\frac{r^2+1}{r}\right)(G_{2}^{(2)})' &- \frac{128(2h-1)h^2}{(r^2+1)^{2h-1}}\left(1+\frac{2}{(r^2+1)}-\frac{1}{(r^2+1)^{2h-1}}\right) \\
	&= 16hG_{N}\Bigg[4h\left(\Omega_{0,0}^{(1)}-h+1\right)\left(f_{0,0}^{(0)}\right)^2+8h\left(h+(h-1)(r^2+1)\right)f_{0,0}^{(0)}f_{0,0}^{(1)} \\
	&-\left(2(r^2+1)-\frac{1}{(r^2+1)^{2h-2}}\right)\left(\partial_{r}f_{0,0}^{(0)}\right)^{2}+2(r^2+1)^2\partial_{r}f_{0,0}^{(0)}\partial_{r}f_{0,0}^{(1)}\Bigg] \,.
\end{split}
\end{equation}
This can be integrated to obtain $G_{2}^{(2)}$ leading to \eqref{eq:G22def}, where the integral of the hypergeometric ${}_{3}{F}_{2}$ can be computed as follows:
\begin{equation}\label{eq:3F2int}
\begin{split}
\int dr\,\frac{r^{3}}{(r^{2}+1)^{2h+2}}\,{}_{3}{F}_{2}\left(1,1,2-2h;2,2;\frac{r^{2}}{r^{2}+1}\right) &= \frac{1}{2}\int dy\, y(1-y)^{2h-1}\,{}_{3}{F}_{2}\left(1,1,2-2h;2,2;y\right) \\
	&= -\frac{1}{2(2h-1)}\int dy\,(1-y)^{2h-1}\sum_{k=1}^{\infty}\frac{(1-2h)_{k}}{k}\frac{y^{k}}{k!} \\
	&= -\frac{1}{2(2h-1)}\sum_{k=1}^{\infty}\frac{(1-2h)_{k}}{kk!}B_{y}(k+1,2h) \,.
\end{split}
\end{equation}
From the way we have written the metric in \eqref{eqn:AAdS3metric}, it is clear that the condition that the metric be smooth at the origin, i.e., that there be no conical singularity at $r=0$, is the condition $G_{2}^{(2)}(r=0) = 0$ which produces \eqref{eq:C3def}.

One can next obtain the correction $G_{1}^{(2)}$ to $g_{tt}$ by solving the $rr$ Einstein's equation at second-order. The term $G_{1}^{(2)}$ is not needed to compute the area, but it is needed to compute the ADM mass in \S\ref{sec:bulkenergy}. The $rr$ Einstein's equation is
\begin{equation}\label{EErr_2ndorder}
\left(R_{rr} - \frac{1}{2}g_{rr}R-g_{rr}\right)\bigg|_{\mathcal{O}(G_{N}^{2})} = 8\pi G_{N}\bra{\psi}T_{rr}\ket{\psi}|_{\mathcal{O}(G_{N})} \,,
\end{equation}
which gives
\begin{equation}\label{eqn:EErr_2ndorder_explicit}
\begin{split}
&\frac{1}{r}\left(\frac{G_{1}^{(2)}}{r^{2}+1}\right)'+2\frac{G_{2}^{(2)}}{(r^{2}+1)^{2}}-64h^{2}\frac{(r^{2}-3)}{(r^{2}+1)^{2h+2}}\left(2-\frac{1}{(r^{2}+1)^{2h-1}}\right) \\ 
	&= 128\pi h\Bigg[2\partial_{r}f_{0,0}^{(0)}\partial_{r}f_{0,0}^{(1)}+\frac{(\Omega_{0,0}^{(0)}f_{0,0}^{(0)})^2}{(r^2+1)^3}+2\frac{\Omega_{0,0}^{(0)}f_{0,0}^{(0)}}{(r^2+1)^2}\left(\Omega_{0,0}^{(1)}f_{0,0}^{(0)}+\Omega_{0,0}^{(0)}f_{0,0}^{(1)}\right)-\frac{8h(h-1)}{(r^2+1)}f_{0,0}^{(0)}f_{0,0}^{(1)} \\
	&+\frac{1}{(r^2+1)^2}\left(1-\frac{1}{(r^2+1)^{2h-1}}\right)\left(\frac{1}{(r^2+1)}\left(\Omega_{0,0}^{(0)}\right)^2-4h(h-1)\right)(f_{0,0}^{(0)})^2\Bigg] \,.
\end{split}
\end{equation}
This can be integrated to obtain $G_{1}^{(2)}$ and one finds
\begin{equation}\label{eq:G12_explicit}
\begin{split}
&G_{1}^{(2)}(r) = \frac{1}{(r^{2}+1)^{4h}}\Bigg[{-}\frac{8h}{(4h+3)}\frac{1}{(r^{2}+1)^{2}}+\frac{h}{(2h+1)}\frac{1}{(r^{2}+1)} \\
	&-\frac{8h^{2}(r^{4}-(2h+3)r^{2}+2h-5)+(r^{2}+1)(2hr^{2}+h+2)}{(4h+1)(4h-1)}\Bigg] \\
	&+\frac{64h^{2}}{(r^{2}+1)^{2h}}\Bigg[\frac{4h^{2}}{(h+1)}\frac{1}{(r^{2}+1)}+\frac{h(2h-1)}{(2h+1)(4h-1)}\left(\psi(2h)-\psi(4h)\right) \\
	&-\frac{2(16h^{3}-12h^{2}+5h-1)}{(2h+1)(4h-1)}-\left(1-2h+\frac{h}{(4h-1)}\left(\psi(2h)-\psi(4h)\right)\right)r^{2}\Bigg] \\
	&-128h^{3}\frac{(2h-1)}{(4h-1)}\Bigg[\frac{r^{4}}{(r^{2}+1)^{2h+1}}{}_{3}{F}_{2}\left(1,1,2-2h;2,2;\frac{r^{2}}{r^{2}+1}\right) \\
	&+(r^{2}+1)\frac{1}{(2h-1)}\sum_{k=1}^{\infty}\frac{(1-2h)_{k}}{kk!}\left(B_{y}(k+1,2h+1)+(1-h+2hy)B_{y}(k+1,2h)+(2h+1)B_{y}(k+2,2h)\right)\Bigg] \\
	&+\mathcal{C}_{3}+\mathcal{C}_{4} \,.
\end{split}
\end{equation}
where the integration constant $\mathcal{C}_{4}$ is fixed by the requirement that the metric be asymptotically AdS, leading to
\begin{equation}\label{eq:C4}
\mathcal{C}_{4} = \frac{64h^{2}}{(2h+1)^{2}(4h-1)}\left(h(1-2h)+(2h+1)(2h^{2}+7h+2)\left(\psi(2h+1)-\psi(4h)\right)\right) \,.
\end{equation}
Thus, we have obtained the full second-order backreacted metric. Note that the $\varphi\varphi$ Einstein's equation does not add any additional constraints to the second-order metric.

%~~~~~~~~~~~~~~~~~~~~~~~~~~~~~~~~~~~~~~~~~~~~~~~
\section{Details of area calculations}
\label{sec:areacalcs}
%~~~~~~~~~~~~~~~~~~~~~~~~~~~~~~~~~~~~~~~~~~~~~~~

In this appendix, we provide some of the details of the area calculations in \S\ref{sec:Area}. For the correction to the area from the change in the metric, we need to compute \eqref{eq:areametricpert_2ndorder}, which we restate here for the reader's convenience:
\begin{equation}\label{eq:areametricpert_2ndorder2}
\bra{\psi}\hat{A}[\Sigma_{A}^{(0)},g]\ket{\psi}\big|_{\mathcal{O}(G_{N}^{2})} = G_{N}^2\int_{r_\mathrm{min}}^{\infty}dr\,\frac{\sqrt{1-\frac{r_{\mathrm{min}}^2}{r^2}}}{(r^2+1)^{\frac{3}{2}}}\left[\frac{64h^{2}}{r^{2}+1}\left(1-\frac{1}{(r^2+1)^{2h-1}}\right)^{2}\left(2+\frac{r_{\mathrm{min}}^{2}}{r^{2}}-r^{2}\right)-2G_{2}^{(2)}(r)\right] \,.
\end{equation} 
The first integral can performed analytically and one finds
\begin{equation}\label{eq:I1}
\begin{split}
\mathcal{I}_{1} &= -\int_{r_\mathrm{min}}^{\infty}dr\,\frac{\sqrt{1-\frac{r_{\mathrm{min}}^2}{r^2}}}{(r^2+1)^{\frac{5}{2}}}\left(64h^2\left(1-\frac{1}{(r^2+1)^{2h-1}}\right)^2\right)\left(r^{2}-2-\frac{r_{\mathrm{min}}^{2}}{r^{2}}\right) \\
	&= -\frac{8}{3}h^2\bigg[8\sin^{2}\left(\frac{\theta}{2}\right)+4\left(\sin^{2}\left(\frac{\theta}{2}\right)+6\theta\cot\left(\frac{\theta}{2}\right)-3\cos^{2}\left(\frac{\theta}{2}\right)-9\right) \\
	&+\sin\theta\csc^{4}\left(\frac{\theta}{2}\right)\left(14\sin\theta-9\theta-6\theta\cos\theta+\sin\theta\cos\theta\right)\bigg] \\
	&-32h^2\frac{\Gamma(\frac{3}{2})\Gamma(2h)}{\Gamma(2h+\frac{7}{2})}\Bigg[{-}\left(2h+\frac{5}{2}\right)\left(4h\cos\theta+3\right)\sin^{4h}\left(\frac{\theta}{2}\right) \\
	&+4h(2h+1)\tan^{4h+4}\left(\frac{\theta}{2}\right)\left(\cot^{2}\left(\frac{\theta}{2}\right)+2\right){}_{2}{F}_{1}\left(2h+2,2h+\frac{5}{2},2h+\frac{7}{2};-\tan^2\left(\frac{\theta}{2}\right)\right) \\
	&+\frac{16h(h+1)(2h+1)}{(7h+4)}\tan^{4h+4}\left(\frac{\theta}{2}\right){}_{2}{F}_{1}\left(2h+\frac{5}{2},2h+3,2h+\frac{9}{2};-\tan^2\left(\frac{\theta}{2}\right)\right)\Bigg] \\
	&-8h^2\frac{\Gamma(\frac{3}{2})\Gamma(4h)}{\Gamma(4h+\frac{5}{2})}\Bigg[16h\left(\cos\theta-2\right)\sin^{8h}\left(\frac{\theta}{2}\right)+\left(\frac{(8h+3)\cot^{2}\left(\frac{\theta}{2}\right)((8h+1)\cot^{2}\left(\frac{\theta}{2}\right)+6)+15}{4h-1}\right)\sin^{8h+2}\left(\frac{\theta}{2}\right) \\
	&\-\frac{64h(4h+1)}{(8h+5)}\tan^{8h+4}\left(\frac{\theta}{2}\right)\csc^{2}\left(\frac{\theta}{2}\right){}_{2}{F}_{1}\left(4h+2,4h+\frac{5}{2},4h+\frac{7}{2};-\tan^2\left(\frac{\theta}{2}\right)\right) \\
	&+\frac{128h(4h+1)(2h+1)}{(8h+5)(8h+7)}\tan^{8h+4}\left(\frac{\theta}{2}\right){}_{2}{F}_{1}\left(4h+\frac{5}{2},4h+3,4h+\frac{7}{2};-\tan^2\left(\frac{\theta}{2}\right)\right)\Bigg] \,.
\end{split}
\end{equation}
This rather complicated result simplifies significantly in the small interval limit to give
\begin{equation}\label{eq:I1_smallint}
\begin{split}
\mathcal{I}_{1} &= h^{2}\theta^{2}\left({-}\frac{16}{3}+\frac{572}{315}\theta^{2}+\mathcal{O}(\theta^4)\right)+32h^2\frac{\Gamma(\frac{3}{2})\Gamma(2h)}{\Gamma(2h+\frac{5}{2})}(4h+3)\left(\frac{\theta}{2}\right)^{4h}\left(1+\mathcal{O}(\theta^2)\right) \\
	&+8h^2\frac{\Gamma(\frac{3}{2})\Gamma(4h-1)}{\Gamma(4h+\frac{5}{2})}\left(\frac{\theta}{2}\right)^{8h-2}\left({-}(8h+1)(8h+3)+(4h-1)\frac{48h+(8h+3)(8h+13)}{3}\left(\frac{\theta}{2}\right)^2+\mathcal{O}(\theta^4)\right) \,.
\end{split}
\end{equation}
We were only able to integrate the second term in the small interval or, equivalently, large $r_{\mathrm{min}}$ limit to obtain
\begin{equation}\label{eq:I2}
\begin{split}
\mathcal{I}_{2} &= -2\int_{r_\mathrm{min}}^{\infty}dr\,\frac{\sqrt{1-\frac{r_{\mathrm{min}}^2}{r^2}}}{(r^2+1)^{\frac{3}{2}}}G_{2}^{(2)}(r) \\
	&= -\frac{64h^{2}}{(4h-1)}\int_{r_\mathrm{min}}^{\infty}dr\,\frac{\sqrt{1-\frac{r_{\mathrm{min}}^2}{r^2}}}{(r^2+1)^{\frac{3}{2}}}\Bigg\{\frac{(4h-1)}{32h^{2}}\mathcal{C}_{3}+\Bigg[{-}\frac{(4h-1)r^{2}+8h-3}{(r^{2}+1)^{4h-1}} \\
	&+\frac{2}{(r^{2}+1)^{2h}}\left((4h-1)(3-4h)r^{2}+4h+2h(r^{2}+1)\left(\psi(2h)-\psi(4h)\right)\right)\Bigg] \\
	&+4h(2h-1)\Bigg[\frac{r^{4}}{(r^{2}+1)^{2h+1}}{}_{3}{F}_{2}\left(1,1,2-2h;2,2;\frac{r^{2}}{r^{2}+1}\right)+\frac{2h}{(2h-1)}\sum_{k=1}^{\infty}\frac{(1-2h)_{k}}{kk!}B_{y}(k+1,2h)\Bigg]\Bigg\} \\
	&= \frac{16h^2}{(4h-1)}\Bigg[\frac{\theta^{2}}{3}\left(1+\frac{1}{60}\theta^2+\mathcal{O}(\theta^4)\right) \\
	&+\frac{\Gamma(\frac{3}{2})\Gamma(2h+1)}{\Gamma(2h+\frac{5}{2})}\left((4h+3)\left((4h-3)(4h-1)-2h\left(\psi(2h)-\psi(4h)\right)\right)\right)\left(\frac{\theta}{2}\right)^{4h}\left(1+\mathcal{O}(\theta^{2})\right) \\
	&+2^{-5}\frac{\Gamma(\frac{3}{2})\Gamma(4h+1)}{\Gamma(4h+\frac{5}{2})}\left(\frac{\theta}{2}\right)^{8h-2}\left(4(8h+1)+\frac{(32h^{2}-52h+17)}{3}\theta^{2}+\mathcal{O}(\theta^{4})\right)\Bigg] \,,
\end{split}
\end{equation}
where one expands the hypergeometric ${}_{3}{F}_{2}$ function and the incomplete Beta function $B_{y}(a,b)$ at large $r$ in order to integrate them. The sum of \eqref{eq:I1_smallint} and \eqref{eq:I2} produces \eqref{eq:areametricpert_2ndorder_final}.

Let us next explain how to calculate the corrections to the area coming from the shift in the surface $\Sigma_{A,\mathrm{geo}}^{(1)}$. This requires computing the integral of the area Langrangian expanded at second order in $G_{N}$ \eqref{eqn:RindlerareaL_2ndorder} evaluated on the solution we found for the shift in the surface $\rho_{\mathrm{geo}}^{(1)}$ in \eqref{eq:shapedeform_soln}. This gives
\begin{equation}\label{eq:area_shapedeform_explicit}
\bra{\psi^{(0)}}\hat{A}[\Sigma_{A,\mathrm{geo}}^{(1)},g]\ket{\psi^{(0)}}|_{\mathcal{O}(G_{N}^{2})} = \frac{1}{2}\int_{-\infty}^{\infty}dx\,\left(({\rho_{\mathrm{geo}}^{(1)}}'(x))^{2}+(\rho_{\mathrm{geo}}^{(1)}(x))^2-V_{1}(x)\rho_{\mathrm{geo}}^{(1)}(x)-V_{2}(x){\rho_{\mathrm{geo}}^{(1)}}'(x)\right) \,.
\end{equation}
All of these integrals are straightforward to compute, except for those involving the hypergeometric function ${}_{2}{F}_{1}$. The $\theta^{2}$ and $\theta^{4}$ terms do not require any such integrals. For the $\theta^{4h}$ type terms, one can compute the integral involving the hypergeometric function using the following series of manipulations:
\begin{equation}\label{eq:hypergeoint}
\begin{split}
\int_{0}^{\infty}z^{2h+k}(1+z)^{-4h-7}&{}_{2}{F}_{1}\left(1,-2h-2,2h+1;-z\right) \\
	&= \int_{0}^{\infty}z^{2h+k}(1+z)^{-4h-8}{}_{2}{F}_{1}\left(1,4h+3,2h+1;\frac{z}{1+z}\right) \\
	&= \int_{0}^{1}du\,u^{2h+k}(1-u)^{2h+6-k}{}_{2}{F}_{1}\left(1,4h+3,2h+1;u\right) \\
	&= \frac{\Gamma(2h+k+1)\Gamma(2h+7-k)}{\Gamma(4h+8)}{}_{3}{F}_{2}\left(1,4h+3,2h+k+1;2h+1,4h+8;1\right) \\
	&= \frac{\Gamma(2h+k+1)\Gamma(2h+7-k)}{\Gamma(4h+8)}\frac{1}{(-2h-k)_{k}} \\
	&\times\sum_{\ell=0}^{k}\frac{(-1)^{\ell}(-2h-k)_{k-\ell}\Gamma(\ell+1)(4h+3)_{\ell}}{(4h+8)_{\ell}}{k\choose \ell}{}_{2}{F}_{1}\left(\ell+1,4h+3+\ell,4h+8+\ell;1\right) \\
	&= \frac{\Gamma(2h+1)\Gamma(2h+k+1)\Gamma(2h+7-k)\Gamma(k+1)}{24\,\Gamma(4h+7)\Gamma(4h+3)}\sum_{\ell=0}^{k}\frac{\Gamma(4h+3+\ell)\Gamma(4-\ell)}{\Gamma(k-\ell+1)\Gamma(2h+1+\ell)} \,,
\end{split}
\end{equation}
where $z \equiv e^{2x}$, $u \equiv \frac{z}{1+z}$, and $k=0,\ldots,3$. In the first line, we used the Pfaff transformation and in the third line, we used 7.512.5 in \cite{GradRyz}. 

The calculation of the $\theta^{8h}$ term appearing in \eqref{eq:area_shapedeform_explicit} turns out to be formidable due to complicated integrals hypergeometric functions, as well as hypergeometric functions squared. We split the calculation into two integrals:
\begin{equation}\label{eq:area_shapedeform_explicit_8h}
\begin{split}
\bra{\psi^{(0)}}\hat{A}[\Sigma_{A,\mathrm{geo}}^{(1)},g]\ket{\psi^{(0)}}|_{\mathcal{O}(G_{N}^{2}),\mathcal{O}(\theta^{8h})} &= \frac{1}{2}\int_{-\infty}^{\infty}dx\,\bigg(({\rho_{\mathrm{geo}}^{(1)}}'(x)|_{\theta^{4h}})^{2}+(\rho_{\mathrm{geo}}^{(1)}(x)|_{\theta^{4h}})^2 \\
	&-V_{1}(x)|_{\theta^{4h}}\rho_{\mathrm{geo}}^{(1)}(x)|_{\theta^{4h}}-V_{2}(x)|_{\theta^{4h}}{\rho_{\mathrm{geo}}^{(1)}}'(x)|_{\theta^{4h}}\bigg) \\
	&\equiv \mathfrak{I}_{1}+\mathfrak{I}_{2} \,,
\end{split}
\end{equation}
where
\begin{equation}\label{eq:frakI}
\begin{split}
\mathfrak{I}_{1} &= \frac{1}{2}\int_{-\infty}^{\infty}dx\,\left(({\rho_{\mathrm{geo}}^{(1)}}'(x)|_{\theta^{4h}})^{2}+(\rho_{\mathrm{geo}}^{(1)}(x)|_{\theta^{4h}})^2\right) \\
\mathfrak{I}_{2} &= -\frac{1}{2}\int_{-\infty}^{\infty}dx\,\left(V_{1}(x)|_{\theta^{4h}}\rho_{\mathrm{geo}}^{(1)}(x)|_{\theta^{4h}}+V_{2}(x)|_{\theta^{4h}}{\rho_{\mathrm{geo}}^{(1)}}'(x)|_{\theta^{4h}}\right) \,.
\end{split}
\end{equation}
The integral of $\mathfrak{I}_{2}$ can be evaluated by following similar manipulations as in the first four lines of \eqref{eq:hypergeoint} and then using Lemma 2.2 in \cite{Lewa} to convert the resulting hypergeometric ${}_{3}{F}_{2}$ functions into ratios of Gamma functions. Once the dust settles, one finds
\begin{equation}\label{eq:frakI2}
\mathfrak{I}_{2} = -\left(\frac{\theta}{2}\right)^{8h}\left(1+\mathcal{O}(\theta^{2})\right)64h^{2}\left(2\frac{\Gamma(\frac{3}{2})^{2}\Gamma(2h+1)^{2}}{\Gamma(2h+\frac{3}{2})^{2}}+\frac{\Gamma(\frac{3}{2})\Gamma(4h+1)}{\Gamma(4h+\frac{5}{2})}\right)\,.
\end{equation}

To compute $\mathfrak{I}_{1}$, we start by rewriting it as
\begin{equation}\label{eq:frakI1}
\begin{split}
\mathfrak{I}_{1} &= \frac{1}{2}\int_{-\infty}^{\infty}dx\,\left(({\rho_{\mathrm{geo}}^{(1)}}'(x)|_{\theta^{4h}})^{2}+(\rho_{\mathrm{geo}}^{(1)}(x)|_{\theta^{4h}})^2\right) \\
	&= \int_{-\infty}^{0}dx\,\left(({\rho^{(1)}}'(x)|_{\theta^{4h}})+(\rho^{(1)}(x)|_{\theta^{4h}})\right)^{2}-\left(\rho^{(1)}(0)|_{\theta^{4h}}\right)^{2} \,,
\end{split}
\end{equation}
where we have used the fact that $\rho(x)$ is an even function of $x$ to restrict the bounds of integration to $-\infty < x < 0$ and we used the boundary condition $\rho^{(1)}(-\infty) = 0$ to set one of the boundary terms to zero. This rewriting is very useful because the sum inside the square simplifies significantly to give
\begin{equation}\label{eq:sumsq}
\begin{split}
&\int_{-\infty}^{0}dx\,\left(({\rho^{(1)}}'(x)|_{\theta^{4h}})+(\rho^{(1)}(x)|_{\theta^{4h}})\right)^{2} \\
	&= \left(\frac{\theta}{2}\right)^{8h}64h^{2}\int_{-\infty}^{0}dx\,\Bigg[2\frac{\Gamma(2h+1)\Gamma(\frac{3}{2})}{\Gamma(2h+\frac{3}{2})}e^{x} \\
	&-\frac{\sech^{4h+1}x}{1+e^{2x}}\left({-}\frac{(h+2)}{(h+1)}+\frac{2h}{(2h+1)}e^{2x}+\frac{1}{(h+1)}{}_{2}{F}_{1}\left(1,-2h-2,2h+1,-e^{2x}\right)\right)\Bigg]^{2} \,.
\end{split}
\end{equation}
The terms not involving hypergeometric functions are straightforward to compute. We will now demonstrate how to compute the integral of the hypergeometric function squared as the other terms involving hypergeometric functions can be computed similarly. 

The strategy is to rewrite the hypergeometric function in terms of the incomplete Beta function and then manipulate the resulting expression using integration by parts and incomplete Beta function identities until one obtains the known integral
\begin{equation}\label{eq:betaint}
\int dv\,v^{a-1}(1-v)^{b-1}B_{v}(a,b) = \frac{1}{2}B_{v}(a,b)^{2} \,.
\end{equation}
One finds
\begin{equation}
\begin{split}
\int_{-\infty}^{0}dx\,&\frac{\sech^{8h+2}x}{(1+e^{2x})^{2}}{}_{2}{F}_{1}\left(1,-2h-2,2h+1,-e^{2x}\right)^{2} \\
	&= 2^{8h+1}\int_{0}^{\infty}dw\,e^{-(4h+1)w}{}_{2}{F}_{1}\left(2h,4h+3,2h+1,-e^{-w}\right)^{2} \\
	&= e^{-4\pi ih}2^{8h+3}h^{2}\int_{-1}^{0}dv\,B_{v}(2h,-4h-2)^{2} \\
	&= e^{-4\pi ih}2^{8h+3}h^{2}\Bigg[\frac{(3h+2)}{2(h+1)}B_{-1}(2h,-4h-2)^{2} - \frac{(h+1)}{2h}B_{-1}(2h+1,-4h-2)^{2} \\
	&+B_{-1}(2h,-4h-2)B_{-1}(2h+1,-4h-2)+\frac{1}{2h}B_{-1}(4h+1,-8h-4)-\frac{1}{2(h+1)}B_{-1}(4h,-8h-4)\Bigg] \\
	&= \frac{1}{2}\bigg((h+1)\frac{\Gamma(\frac{3}{2})^{2}\Gamma(2h+1)^{2}}{\Gamma(2h+\frac{3}{2})^{2}}+2(h+1)^{2}(2h+1)\frac{\Gamma(\frac{3}{2})\Gamma(2h+1)}{\Gamma(2h+\frac{3}{2})}+\frac{(4h^{3}+13h^{2}+10h+2)}{2(4h+1)(2h+1)^{2}} \\
	&-(4h+3)h\frac{\Gamma(\frac{3}{2})\Gamma(4h+1)}{\Gamma(4h+\frac{5}{2})}\bigg) \,,
\end{split}
\end{equation}
where $w=-2x$, $v=-e^{-w}$, and in the second line we used the Euler transformation. The $B_{-1}$ functions are converted into Gamma functions on the last line by writing $B_{-1}$ in terms of hypergeometric functions and using known expressions for these with final argument evaluated at $-1$. Performing the integration of the other terms in \eqref{eq:sumsq} in a similar manner leads to the final result
\begin{equation}\label{eq:frakI1final}
\mathfrak{I}_{1} = \left(\frac{\theta}{2}\right)^{8h}\left(1+\mathcal{O}(\theta^{2})\right)32h^{2}\left(2\frac{\Gamma(\frac{3}{2})^{2}\Gamma(2h+1)^{2}}{\Gamma(2h+\frac{3}{2})^{2}}+\frac{\Gamma(\frac{3}{2})\Gamma(4h+1)}{\Gamma(4h+\frac{5}{2})}\right) \,.
\end{equation}
Therefore, the $\theta^{8h}$ contribution to the shape deformation of the area is given by
\begin{equation}\label{eq:area_shapedeform_explicit_8h_final}
\begin{split}
\bra{\psi^{(0)}}\hat{A}[\Sigma_{A,\mathrm{geo}}^{(1)},g]\ket{\psi^{(0)}}&|_{\mathcal{O}(G_{N}^{2}),\mathcal{O}(\theta^{8h})} \\
	&= -\left(\frac{\theta}{2}\right)^{8h}\left(1+\mathcal{O}(\theta^{2})\right)32h^{2}\left(2\frac{\Gamma(\frac{3}{2})^{2}\Gamma(2h+1)^{2}}{\Gamma(2h+\frac{3}{2})^{2}}+\frac{\Gamma(\frac{3}{2})\Gamma(4h+1)}{\Gamma(4h+\frac{5}{2})}\right) \,.
\end{split}
\end{equation}
%

%~~~~~~~~~~~~~~~~~~~~~~~~~~~~~~~~~~~~~~~~~~~~~~~
%\section{Komar mass}
%\label{sec:Komarmass}
%~~~~~~~~~~~~~~~~~~~~~~~~~~~~~~~~~~~~~~~~~~~~~~~

%%%%%%%%%%%%%%%%%%%%%%%%%%%%%%%%%%%%%%%%%%%%%%%%%%%%%%%

\bibliography{qes1-refs} 
\bibliographystyle{JHEP}

%%%%%%%%%%%%%%%%%%%%%%%%%%%%%%%%%%%%%%%%%%%%%%%%%%%%%%%

\end{document}